\documentclass[11pt,a4paper]{article}
\pdfoutput=1
\usepackage{jstyle}

\newcommand{\bi}{\begin{itemize}}
\newcommand{\ei}{\end{itemize}}
\newcommand{\bea}{\begin{eqnarray}}
\newcommand{\eea}{\end{eqnarray}}
\newcommand{\rmd} {{\rm d}}

\author{Seungho GWAK}
\author{\quad Euihun JOUNG}
\author{\quad Karapet MKRTCHYAN}
\author{\quad Soo-Jong REY}

\affiliation{School of Physics \& Astronomy and Center for Theoretical Physics \\ Seoul National University, Seoul 08826 \rm KOREA}
\affiliation{Gauge, Gravity \& Strings, \ Center for Theoretical Physics of the Universe\\
Institute for Basic Sciences, Daejeon 34047 \rm KOREA}

\emailAdd{epochmaker, 
\ euihun.joung, \ karapet, \ sjrey @ snu.ac.kr}

\title{\centering
\LARGE{Rainbow Valley \\
of \\ Colored (Anti) de Sitter Gravity in Three Dimensions}}

\abstract{We propose a theory of three-dimensional (anti) de Sitter gravity carrying Chan-Paton color charges. We define the theory by Chern-Simons formulation with the gauge algebra $(\mathfrak{gl}_{2}\oplus \mathfrak{gl}_{2})\otimes \mathfrak{u}(N)$, obtaining a color-decorated version of interacting spin-one and spin-two fields. We also describe the theory in metric formulation and show that, among $N^{2}$ massless spin-two fields, only the singlet one plays the role of metric graviton whereas the rest behave as \emph{colored spinning matter} that strongly interacts at large $N$.
Remarkably, these \emph{colored spinning matter} acts as Higgs field and generates a non-trivial potential of staircase shape. At each extremum labelled by $k = 0, \ldots, 
[\frac{N-1}2]$, the $\mathfrak{u}(N)$ color gauge symmetry is spontaneously broken down to $\mathfrak{u}(N-k)\oplus \mathfrak{u}(k)$ and provides different (A)dS backgrounds with the cosmological constants  $\big(\frac{N}{N-2k}\big)^{2}\,\L$\,.
When this symmetry breaking takes place, the spin-two Goldstone modes combine with (or are eaten by) the spin-one gauge fields to become partially-massless spin-two fields. We discuss various aspects of this theory  and highlight physical implications.
}

\begin{document}

%\date{\currenttime}

\maketitle

\rightline{\sl``It's time to try. Defying gravity.}
\rightline{\sl I think I'll try. Defying gravity.}
\rightline{\sl 
And you can't pull me down!"}
\rightline{--- Gregory Maguire}
\rightline{`Wicked: The Life and Times of the Wicked Witch of the West'} 
\vskip0.3cm

\section{Introduction}

The Einstein's theory of gravity is known to be rigid. Variety of modifications has been challenged  
with diverse motivations, yet no concrete result of success has been reported so far (for related readings, see e.g. \cite{modified gravity} and references therein).  Recently, two situations defying the rigidity of Einstein gravity were actively explored. One is the massive modification of gravity 
\cite{massive}, along with numerous variants in three dimensions \cite{3d}. Another is higher-derivative modifications of the gravity \cite{HD}.

In this work, we investigate the modification of  Einstein gravity to a multi-graviton theory: \emph{the color decoration}. In spite of previous negative results \cite{Wald, Boulanger:2000rq},
certain models of colored gravity can be consistently constructed
by introducing other field contents than massless spin-two fields only.
Moreover, 
the color-decoration we study is not limited to the Einstein gravity and can be applied to various extensions of it. In particular, all higher-spin theories formulated in \cite{Vasiliev} can be straightforwardly color-decorated, whose first steps were conceived in \cite{Color Vasiliev}.
 In the companion paper \cite{Gwak:2015jdo}, we study a three-dimensional color-decorated higher-spin gravity.
 
The color decoration of gravity evokes various conceptual issues. 
Clearly, the colored gravity is analogous to Yang-Mills theory 
were if the Einstein gravity compared to Maxwell theory. Besides
the presence of multiple gauge bosons in the system, the Yang-Mills theory as color-decorated Maxwell theory has  far-reaching 
consequences that are not shared by the Maxwell theory.\footnote{The story goes that, during C.N. Yang's seminar at the Institute for Advanced Study at Princeton in 1953, Wolfgang Pauli commented that he first discovered non-Abelian gauge theory in this manner, but then immediately dismissed it because vector bosons are massless and hence ``unphysical". We acknowledge Stanley Deser for straightening us up for details of this history.}  Likewise, we anticipate that color-decorated gravity brings out surprising new features one could not simply guess on a first look. 
In this paper, we define and study a version of the color-decorated Einstein gravity in three dimensions, and uncover remarkable new features not shared by the Einstein gravity itself. Most interestingly, we will find that this color-decorated gravity admits a number of  (A)dS backgrounds with different  cosmological constants as classical vacua.

In anaylzing our model of three-dimensional color-decorated gravity, we shall make use of  both the Chern-Simons formulation \cite{Achucarro:1987vz} and the metric formulation. Various features of the theory are more transparent in one formulation over the other. For instance, the existence of multiple (A)dS vacua with different cosmological constants can be understood more intuitively in the metric formulation,  whereas consistency of the theory is more manifest in the Chern-Simons formulation. The latter makes use of the gauge algebra, 
\be
\mathfrak{g} \, = \, 	(\mathfrak{gl}_{2}\oplus\mathfrak{gl}_{2})\otimes \mathfrak{u}(N)\,,
\ee
where the $\mathfrak{u}(N)$ and $\mathfrak{gl}_{2}\oplus\mathfrak{gl}_{2}$
correspond respectively to the color gauge algebra and the extended isometry algebra governing the gravitational dynamics.  
We stress that, compared to the usual gravity with $\mathfrak{sl}_{2}\oplus\mathfrak{sl}_{2}$ gauge algebra,
the color-decorated gravity has two additional identity generators from each of $\mathfrak{gl}_{2}$\,. They are indispensable for the consistency of color decoration and correspond to two additional Chern-Simons gauge fields on top of the graviton.
Hence, when colored-decorated, we get a massless spin-two field and two 
non-Abelian spin-one fields, both taking adjoint values of $\mathfrak{u}(N)$.
Let us also remark that compared to the spin-one situation where the Abelian Maxwell theory turns into the non-Abelian Yang-Mills theory once color-decorated, Einstein gravity is already non-Abelian, while color decoration enlarges the gauge algebra of the theory.

Re-expressing the theory in metric formulation makes it clear that, among $N^{2}$ massless spin-two fields, only the singlet one plays the role
of genuine graviton, viz. the first fundamental form, whereas the rest rather behave as \emph{colored spinning matter} fields with minimal covariant coupling to the gravity as well as to the $\mathfrak{u}(N)$ gauge fields.
We derive the explicit form of Lagrangian for these \emph{colored spinning matter} fields 
and find that they have a strong self-coupling compared to the gravitational one by the factor of $\sqrt{N}$\,.
Analyzing the potential of the Lagrangian, we identify
all the extrema:
there are $[\frac{N+1}2]$ number of them and they have 
different  cosmological constants,
\be
	\left(\frac{N}{N-2k}\right)^{\!2}\,\L\,,
	\label{L k}
\ee
where $k=0,\ldots, [\frac{N-1}2]$ is the label of the extrema and $\L$ is the cosmological constant of the vacuum
with maximum radius (corresponding to $k=0$).
Note that not only (A)dS but also any exact gravitational backgrounds such as BTZ black holes \cite{Banados:1992wn} lie multiple times with different cosmological constants \eqref{L k} 
in the vacua of the colored gravity.
All extrema except the $k=0$ vacuum spontaneously break the color symmetry $U(N)$ down to $U(N-k)\times U(k)$\,.
When this symmetry breaking takes place,
the corresponding $2\,k\,(N-k)$ spin-two Goldstone modes are combined with the gauge fields to become 
 the partially-massless spin-two fields \cite{Deser-Waldron}:
 the latter spectrum does not have any propagating degrees of freedom (DoF) similarly to the massless ones.
Instead in AdS case, they describe `four' boundary DoF which originate from the boundary modes of the colored massless spin-two and spin-one fields.

The organization of the paper is as follows. 
In Section \ref{sec: nogo}, we recapitulate the no-go theorem of interacting theory of multiple massless spin-two fields. In Section \ref{sec: CS}, we define the color-decorated (A)dS$_3$ gravity in Chern-Simons formulation, and discuss how this theory evades the no-go theorem. In Section \ref{sec: metric}, we recast the Chern-Simons action
into metric formulation by solving torsion condition
and obtain the Lagrangian for the colored massless spin-two fields. 
In Section \ref{sec: vacua}, we solve the equations of motion and find a class of classical vacua with varying degrees of color symmetry breaking. We show that these (A)dS vacua have different cosmological constants. We explicitly investigate the simplest
example of  $k=1$ vacuum in $N=3$ case. 
In Section \ref{sec: rainbow}, we expand the theory around a color non-singlet vacuum and analyze the field spectrum contents. We demonstrate that the fields corresponding to the broken part of the color symmetry describe the spectrum of partially-massless spin-two field.
Section \ref{sec: discussion} contains discussions of our results and outlooks.
Finally, Appendix \ref{sec: PM} reviews massive and (partially-)massless spin-two fields in three-dimensions.

%%%%%%%%%%%%%%%%%%%%%%%%%%%%%%%%%%%%%%%%%%%%%%%%%%%%%%%%%

\section{No-Go Theorem on Multiple Spin-Two Theory}
\label{sec: nogo}

Einstein gravity describes the dynamics of massless spin-two field on a chosen vacuum.  
Conversely, it can also be verified that the Einstein gravity is the only 
interacting theory of a massless spin-two field (see e.g. \cite{GR}).
In this context, one may ask whether there exists 
a non-trivial theory of multiple massless spin two fields.
This possibility has been examined in \cite{Wald,Boulanger:2000rq}, leading to a no-go theorem. We shall begin our discussion by reviewing this result.\footnote{See also related discussion in \cite{BA}.}

The no-go theorem asserts that there is no interacting theory 
of multiple massless spin-two fields, without inclusion of other fields.  
The first point to note in this consideration is that any gauge-invariant two-derivative cubic interactions among 
the spin-two fields is in fact equivalent to that of Einstein-Hilbert (EH) action, 
modulo \emph{color-decorated} cubic coupling constants $g_{IJK}$\,:
\be
	g_{IJK}\left(h^{I}_{\m\r}\,\partial^{\r}\,h^{J}_{\n\l} \partial^{\l} h^{K\,\m\n}+\cdots\right).
\ee
Here, $h_{\m\n}^{I}$ are the massless spin-two fields with \emph{color} index 
$I$\,, and the tensor structure inside of the bracket is  that of the EH cubic vertex. 
For the consistency with the color indices, it is required that 
the coupling constants are fully symmetric: $g_{IJK}=g_{(IJK)}$\,. Moreover,
the gauge invariance requires that these constants define a Lie algebra 
spanned by the colored isometry generators. For instance, in the Minkowski spacetime, the colored generators
$P^{I}_{\m}$ and $M^{I}_{\m\n}$ obey
\be
	[\,M_{\m\n}^{I}\,,\,P^{J}_{\r}\,]=2\,g^{IJ}{}_{K}\,\eta_{\r[\n}\,P^{K}_{\m]}\,,
	\qquad
	[\,M_{\m\n}^{I}\,,\,M^{J}_{\r\l}\,]
	=4\,g^{IJ}{}_{K}\,\eta_{[\n[\r}\,M^{K}_{\l]\m]}\,.
\ee
Relating these colored generators to the usual isometry ones as
 $P^{I}_{\m}=P_{\m}\otimes\bm T^{I}$
and $M^{I}_{\m\n}=M_{\m\n}\otimes\bm T^{I}$\,,
one can straightforwardly conclude that
the color algebra $\mathfrak{g}_{c}$
generated by $\bm T^{I}$ must be
 \emph{commutative} and \emph{associative} \cite{Wald}.
Moreover, one can even show that $\mathfrak{g}_{c}$ necessarily
reduces to a direct sum of one-dimensional ideals \cite{Boulanger:2000rq}:
$\bm T^{I}\,\bm T^{J}=0$ for $I\neq J$\,.
Therefore, in this set-up, the only possibility is the simple sum of 
several copies of Einstein gravity which do not interact with each other.

This no-go theorem can be evaded with a slight generalization of the setup.
Firstly, if the isometry algebra can be consistently extended from a Lie algebra to an associative one,
then the commutativity condition on the color algebra $\mathfrak{g}_{c}$ can be relaxed. The associative extension of isometry algebra 
typically requires to include other spectra,  such as spin-one and possibly higher spins \cite{Color Vasiliev}.
Moreover, it is not necessary to require that the structure constants  $g_{IJK}$ of  $\mathfrak{g}_c$ be totally symmetric, but 
sufficient to assume that the totally symmetric part is non-vanishing, $g_{(IJK)}\neq 0$\,, so that massless spin-two fields have non-trivial interactions among themselves. 

Hence, an interacting theory of multiple massless spin-two fields might be viable once other fields are added and coupled to them.
As the next consistency check, one can examine the fate of
the general covariance in such a theory:
if there exists a genuine metric field
among these massless spin-two fields, 
the others should be subject to interact covariantly with gravity.
Moreover, one can also examine whether the multiple massless spin two fields
can be color-decorated bona fide by carrying non-Abelian charges. 
In principle, a theory can be made to covariantly interact 
with gravity or non-Abelian gauge field 
by simply replacing all its derivatives
by the covariant ones with respect to both the diffeomorphism transformation
and the non-Abelian gauge transformation.
However, as in the diffeomorphism-covariant interactions
of higher-spin fields, such replacements  spoil
the gauge invariance of the original system  \cite{AD}.
The problematic term in the gauge variation is 
proportional to the curvatures, namely, Riemann tensor $R_{\m\n\r\l}$
or non-Abelian gauge field strength $F_{\m\n}$\,. In three-dimensions, fortuitously, this is not a problem
as these curvatures are just proportional to the field equations of Eintein gravity or Chern-Simons gauge theory, respectively.
In higher dimensions, these terms can be compensated  by introducing a non-trivial cosmological constant,
but at the price of adding higher-derivative interactions \cite{FV,Joung:2013nma}.

So, we conclude that, to have a consistent interacting theory of color-decorated massless spin-two fields, 
we need an (A)dS isometry gauge algebra which can be extended to an associative one. 
An immediate candidate is higher-spin algebra in any dimensions,
since Vasiliev's higher-spin theory can be consistently color-decorated, as mentioned before.
Other option is to take the isometry algebras of $(A)dS_3$ and $(A)dS_5$ 
which are isomorphic to $\mathfrak{sl}_{2}\oplus\mathfrak{sl}_{2}$
and $\mathfrak{sl}_{4}$\, and can be extended to associative ones,
$\mathfrak{gl}_{2}\oplus\mathfrak{gl}_{2}$
and $\mathfrak{gl}_{4}$ by simply adding unit elements corresponding to spin-one fields.

%%%%%%%%%%%%%%%%%%%%%%%%%%%%%%%%%%%%%%%%%%%%%%%%%%%%%%%%%%%%%%%%%%%%%%%
\section{Color-Decorated (A)dS$_3$ Gravity: Chern-Simons Formulation}
\label{sec: CS}

Let us now move to the explicit construction of a theory of colored gravity. 
In this paper, we focus on the case of three-dimensional gravity.

%%%%%%%%%%%%%%%%%%%%%%%%%%%%%%%%%%%%%%%%%%%%%%%%%%%%%%%%
\subsection{Color-Decorated Chern-Simons Gravity}
In the uncolored case, it is known that the three-dimensional gravity can be formulated as a Chern-Simons theory with the action
\be
S[{\cal A}] =\frac{\k}{4\pi}\, \int 
\tr
\Big(
\cA\wedge d\cA
+\frac{2}{3}\,\cA\wedge \cA\wedge \cA
\Big)\,,
\label{CS}
\ee
for the gauge algebra $\mathfrak{sl}_{2}\oplus\mathfrak{sl}_{2}$\,. 
The constant $\k$ is the level of Chern-Simons action.
We are interested in color-decorating this theory. 
Physically, this can be done by attaching Chan-Paton factors to the 
gravitons. Mathematically, this amounts to requiring the fields to take values in 
the tensor-product space $\mathfrak{g}_i \otimes \mathfrak g_c$\,,
where the  $\mathfrak{g}_{i}$ is the isometry part of the algebra including 
$\mathfrak{sl}_{2}\oplus\mathfrak{sl}_{2}$
and the
$\mathfrak{g}_{c}$ is a finite-dimensional Lie algebra of a matrix group 
\mt{\mathfrak{G}_c}\,. 
For generic Lie algebras $\mathfrak g_i$ and $\mathfrak g_c$\,,
their tensor product do not form a Lie algebra,
as is clear from the 
commutation relations:
\be
	[\,M_{X}\otimes \bm T_{I},M_{Y}\otimes \bm T_{J}\,]
	=\frac12\,[\,M_{X},M_{Y}\,]\otimes \{\,\bm T_{I} ,\bm T_{J}\,\}
	+\frac12\,\{\,M_{X}, M_{Y}\,\} \otimes [\,\bm T_{I},\bm T_{J}\, ] \,.
\label{commutation of product group}
\ee
The anticommutators $\{\bm T_{I},\bm T_{J}\}$ and $\{M_{X}, M_{Y}\}$ do not make sense within the Lie algebras.
Instead, if we start from associative algebras $\mathfrak{g}_i$ and $\mathfrak{g}_c$\,, their direct product $\mathfrak g_i \otimes \mathfrak g_c$ will form an associative algebra, from which we can also obtain the Lie algebra structure.
Hence, in this paper, we will consider associative algebras for $\mathfrak{g}_{i}$ and $\mathfrak{g}_{c}$\,.
For the color algebra $\mathfrak{g}_{c}$\,, we take the matrix algebra 
$\mathfrak{u}(N)$. For the isometry algebra $\mathfrak{g}_{i}$\,,
we take $\mathfrak g_i = \mathfrak{gl}_{2} \oplus \mathfrak{gl}_2$ 
(instead of $\mathfrak{sl}_{2} \oplus \mathfrak{sl}_2$).
The trace $\tr$ of \eqref{CS} should be defined also in the 
tensor product space and is given by the product of two traces as 
\be
	\tr(\mathfrak{g}_{i}\otimes \mathfrak{g}_{c}) :=
	\tr(\mathfrak{g}_{i})\,\tr(\mathfrak{g}_{c})\,.
\ee
We also need for the fields to obey Hermicity conditions
compatible with the real form of the complex algebra.\footnote{Note that if the isometry algebra $\mathfrak g_i$ is not associative --- as is the case with Poincar\'e algebra discussed in \cite{Wald,Boulanger:2000rq} --- then the requirement of the closure of the algebra is that the color algebra $\mathfrak g_c$ be associative (for the first term in \eqref{commutation of product group} to be in the product algebra) and commutative (for the second term in \eqref{commutation of product group} to vanish).}

Therefore, our model of colored gravity is the Chern-Simons theory \eqref{CS}
where the one-form gauge field $\cA$ takes value in 
\be
	\mathfrak{g}=\left(\mathfrak{gl}_{2}\oplus \mathfrak{gl}_{2}\right)
	\otimes \mathfrak{u}(N)
	\ \ominus\ 
	{\rm id}
	\otimes \bm I\,.
	\label{g}
\ee
Notice that we have subtracted the 
${\rm id}\otimes \bm I$,  where $\rm id$ and $\bm I$ are the centers
of $\mathfrak{gl}_{2}\oplus \mathfrak{gl}_{2}$ and $\mathfrak{u}(N)$\,, 
respectively: it corresponds to an Abelian vector field (described by Chern-Simons action) which does not interact with other fields.\footnote{In the Introduction, we sketched our model without taking into account this subtraction for the sake of simplicity.} 
As a complex Lie algebra, $\mathfrak{g}$ in \eqref{g} is in fact isomorphic
to $\mathfrak{sl}_{2N}\oplus \mathfrak{sl}_{2N}$\,.
This can be understood from the fact that
the tensor product of $2\times 2$ and $N\times N$ matrices
gives $2N\times 2N$ matrix.
It would be worth to remark as well that 
the algebra $\mathfrak{g}$ necessarily contains 
elements in ${\rm id}\otimes \mathfrak{su}(N)$
which correspond to 
the gauge symmetries of $\mathfrak{su}(N)$ Chern-Simons theory.
 In this sense, this $\mathfrak{su}(N)$
will be referred to as the color algebra.

It turns out useful\footnote{Later, we will take advantage of this decomposition in solving the torsionless condition to convert Chern-Simons formulation into metric formulation.} 
to decompose the algebra $\mathfrak{g}$ \eqref{g} 
into two orthogonal parts as
\be
	\mathfrak{g}=\mathfrak{b}\oplus\mathfrak{c}\,,
	\qquad \mbox{such that} \qquad
	\tr(\mathfrak{b}\,\mathfrak{c})=0\,,
	\label{g=b+c}
\ee
where $\mathfrak{b}$ is the subalgebra:
\be
	[\mathfrak{b},\mathfrak{b}]\subset\mathfrak{b}\,,
\ee
corresponding to the \emph{gravity plus gauge} sector (mediating gravity and gauge forces),
whereas $\mathfrak{c}$ corresponds to the \emph{matter} sector
--- including all colored spin-two fields ---
subject to the covariant transformation,
\be
	[\mathfrak{b},\mathfrak{c}]\subset \mathfrak{c}\,.
\ee
Corresponding to the decomposition \eqref{g=b+c}, 
the one-form gauge field $\cal A$ can be written as
the sum of two parts
\be
	\cA=\cB+\cC\,,
\ee
where $\cB$ and $\cC$ takes value in $\mathfrak{b}$ and $\mathfrak{c}$\,, respectively.
In terms of $\cB$ and $\cC$\,,
the Chern-Simons action \eqref{CS} is reduced to
\be
	S[ {\cal B}, {\cal C}] =
	\frac{\k}{4\pi}\, \int \tr
	\left(
	\cB\wedge d\,\cB
	+\frac{2}{3}\,\cB\wedge \cB\wedge \cB
	+\cC\wedge D_{\cB}\,\cC
	+\frac{2}{3}\,\cC\wedge \cC\wedge \cC
	\right),
\ee
where $D_\cB$ is the the $\cB$-covariant derivative:
\be
	D_{\cB}\,\cC=
	d\,\cC+
	\cB\wedge\cC
	+\cC\wedge\cB\,.
\ee
This splitting will prove to be a useful guideline
in keeping manifest covariance with respect to the 
diffeomorphism and the non-Abelian gauge transformation.

%%%%%%%%%%%%%%%%%%%%%%%%%%%%%%%%%%%%%%%%%%%%%%%%%%%%%%%%%
\subsection{Basis of  Algebra}
For further detailed analysis, we set our conventions and notations of the associative algebra involved. The $\mathfrak{sl}_{2}$ has three generators $J_0, J_1, J_2$. Combining them with the center generator $J$\,, one obtains $\mathfrak{gl}_{2}={\rm Span}\{J, J_0, J_1, J_2 \}\,$ with the product,
\be
	J_{a}\,J_{b}=\eta_{ab}\,J+\e_{abc}\,J^{c}\, \qquad \qquad [\,a,b,c =0, 1, 2\,]\,.
	\label{gl2}			
\ee
The $\eta_{ab}$ is the flat metric with mostly positive signs
and $\epsilon_{abc}$ is the Levi-civita tensor of $\frak{sl}_2$ with sign convention $\epsilon_{012}=+1$\,.
The generators of the other $\mathfrak{gl}_{2}$ will be denoted by
$\tilde J_{a}$ and $\tilde J$\,.
In the case of AdS$_3$ background, the real form of the isometry algebra corresponds to
$\mathfrak{so}(2,2)\simeq\mathfrak{sl}(2,\mathbb R)\oplus 
\mathfrak{sl}(2,\mathbb R)$, which satisfy
\be
	(J_{a}, \tilde J_{a})^{\dagger}=-
	(J_{a},\tilde J_{a})\,,
	\qquad (J,\tilde J\,)^{\dagger}= (J,\tilde J\,)\,. 
	\label{AdS J}
\ee
In the case of dS$_3$ background, the real form of the isometry algebra corresponds to  
$\mathfrak{so}(1,3)\simeq\mathfrak{sl}(2,\mathbb C)$, which satisfy
\be
	(J_{a}, \tilde J_{a})^{\dagger}=-
	(\tilde J_{a},J_{a})\,,
	\qquad (J,\tilde J\,)^{\dagger}= (\tilde J,J\,)\,.
	\label{dS J}
\ee
Defining the Lorentz generator $M_{ab}$ and the translation generator $P_{a}$ as
\be
	M_{ab}=\frac12\,\epsilon_{ab}{}^{c}\,\big(J_{c}+\tilde J_{c}\big)\,,
	\qquad
	P_{a}=\frac{1}{2\sqrt{\s}}\,\big(J_{a}-\tilde J_{a}\big)\,,
	\label{MP J}	
\ee
where $\s=+1$ for AdS$_3$ and $\s=-1$ for dS$_3$,
we recover the standard commutation relations
\be
\left[\,M_{ab}, M_{cd}\,\right]= 2\left(\eta_{d[a}\,M_{b]c}
+\eta_{c[b}\,M_{a]d}\right),\quad
\left[\,M_{ab},P_c\,\right]=2\,\eta_{c[b}\,P_{a]}\,				
,\quad 
\left[\,P_a,P_b\,\right]= \s\,M_{ab}\,,
\ee
of $\mathfrak{so}(2,2)$ and $\mathfrak{so}(1,3)$
for $\s=+1$ and $-1$, respectively.
The reality structure of $\mathfrak{gl}_2$  determines
that of the full algebra $\mathfrak{g}$ in \eqref{g}.
As we remarked before, the latter is isomorphic to 
$\mathfrak{sl}_{2N}\oplus \mathfrak{sl}_{2N}$\,,
hence the conditions \eqref{AdS J} and \eqref{dS J} define
which real form of $\mathfrak{sl}_{2N}$ we are dealing with.

The color algebra $\mathfrak{su}(N)$ can be supplemented with the center $\bm I$ to form the associative algebra $\mathfrak{u}(N)$\,, with the product
\be
	\bm{T}_I\,\bm{T}_J=
	\frac{1}{N}\,\delta_{IJ} {\bm I}
	+\left(g_{IJ}{}^{K}+i\,f_{IJ}{}^{K}\right)\bm{T}_K
	\label{u(N)} \qquad
	[I, J, K = 1, \ldots, N^2 - 1]\,.
\ee
The totally symmetric and anti-symmetric structure constants $g_{IJK}$ and $f_{IJK}$ are both real-valued.

We normalize the center generators of both algebras such that their traces are given by\footnote{We use the same notation $\tr$ for the traces of both the isometry algebra and the color algebra.} 
\be
	\tr(J)=2\sqrt{\s}, \qquad \tr(\tilde J) =- 2 \sqrt{\s}, 
	\qquad
	\tr(\bm{I})=N\, . 
\ee
The traces of all other elements vanish. This also defines the trace convention in the Chern-Simons action (\ref{CS}). With the associative product defined in \eqref{gl2}\,, these traces yield all the invariant multilinear forms. For instance, we get the bilinear forms,
\be
	\tr(J_a\,J_b)=2\,\sqrt{\s}\,\eta_{ab}
\, , \qquad 
\tr(\tilde J_{a}\,\tilde J_{b}) = - 2 \, \sqrt{\s} \, \eta_{ab} \,,
	\qquad 
	\tr(\,\bm{T}_I\,\bm{T}_J)=\delta_{IJ}\,,
\ee
which extract the quadratic part of the action.

%%%%%%%%%%%%%%%%%%%%%%%%%%%%%%%%%%%%%%%%%%%%%%%%%%%%%%%%%

In the Chern-Simons formulation, the equation of motion is the zero curvature condition: $\cF=0\,.$ In searching for classical solutions, we choose to decompose the subspaces $\mathfrak{b}$
and $\mathfrak{c}$ in \eqref{g=b+c} as
\be
	\mathfrak{b}=\mathfrak{b}_{\rm GR}\oplus
	\mathfrak{b}_{\rm Gauge}\,,
	\qquad
	\mathfrak{c}= \mathfrak{iso}\otimes \mathfrak{su}(N)\,.
	\label{BC decomp}
\ee
Here, the gravity plus gauge sector corresponds to
\be
	\mathfrak{b}_{\rm GR}=\mathfrak{iso}\otimes \bm I\,,
	\qquad
	\mathfrak{b}_{\rm Gauge}=
	{\rm id}\otimes \mathfrak{su}(N)\, , 
	\label{b content}
\ee
in which $\mathfrak{iso}$ stands for the isometry algebra of the (A)dS$_{3}$ space:
\be
	\mathfrak{iso}=\mathfrak{sl}_{2}\oplus \mathfrak{sl}_{2}\,.
	\label{left right}
\ee

There is a trivial vacuum solution where the connection $\cA$ is nonzero only for the color-singlet component:
\be
	\cB =\left(\frac12\,\o^{ab}\,M_{ab}+\frac1\ell\,e^{a}\,P_{a}\right)\bm I\,, \qquad \cC = 0\,. 
\ee
The zero-curvature condition imposes to $\o^{ab}$ and $e^{a}$ the usual zero (A)dS curvature and zero torsion conditions:
\ba
	&&d\,\omega^{ab}
	+\omega^{a}{}_{c}\wedge\omega^{cb}
	+\frac{\s}{\ell^{2}}\,e^{a}\wedge e^{b}=0\,,\\
	&&d\,e^a+\omega^{ab}\wedge e_b=0\,,
	\label{Torsion}
\ea
which define the (A)dS$_3$ space with the radius $\ell$, or equivalently with the cosmological constant $\L=-(\s/\ell^{2})$.

For a general solution, we again decompose  
$\cA=\cB+\cC$ according to \eqref{BC decomp}.
The gravity plus gauge sector takes the form
\be
	\cB=\left[\frac12\left(
	\o^{ab}+\frac1\ell\,\O^{ab}\right) M_{ab}+\frac1\ell\,e^{a}\,P_{a}\right]\bm I
	+\bm A+\tilde{\bm A}\,, 
	\label{B detail}
\ee
where $\bm A=A^{I}\,J\,\bm T_{I}$ and 
$\tilde{\bm A}=\tilde A^{I}\,\tilde J\,\bm T_{I}$
are two copies of $\mathfrak{su}(N)$ gauge field with
\be
	(\bm A,\tilde {\bm A})^{\dagger}=-\left\{
	\begin{array}{c}
	(\bm A,\tilde {\bm A})\qquad  [\s=+1] \vspace{5pt}\\
	(\tilde {\bm A},\bm A)\qquad  [\s=-1]
	\end{array}\right..
	\label{A herm}
\ee
In \eqref{B detail}, the splitting $\o^{ab}+{ 1\over \ell} \O^{ab}$ in the gravity part is arbitrary and is purely for later convenience. The matter sector is composed of
\be
	\cC=\frac1{\ell}\left(\bm\varphi^{a}\,J_{a}
	+\tilde{\bm\varphi}^{a}\,\tilde J_{a}\right). 
	\label{C detail}
\ee
Here, the colored massless spin-two fields 
$\bm \varphi^{a}=\varphi^{a,I}\,\bm T_{I}$
and $\tilde{\bm\varphi}^{a}=\tilde\varphi^{a,I}\,\bm T_{I}$
take value in $\mathfrak{su}(N)$ carrying 
the adjoint representation. They satisfy 
\be
	(\bm\varphi^{a},\tilde {\bm \varphi}^{a})^{\dagger}=+\left\{
	\begin{array}{c}
	(\bm\varphi^{a},\tilde {\bm \varphi}^{a})\qquad  [\s=+1] \vspace{5pt}\\
	(\tilde {\bm \varphi}^{a},\bm\varphi^{a})\qquad  [\s=-1]
	\end{array}\right..
	\label{varphi dagger}
\ee
Note that the above has a sign difference from \eqref{A herm}.

We may find solutions by demanding that (\ref{B detail}) and (\ref{C detail}) solve for the zero curvature condition. While this procedure straightfowardly yields nontrivial solutions, for better physical interpretations, we shall first recast the Chern-Simons formulation to the metric formulation and then obtain these nontrivial solutions by solving the latter's field equations.    

%%%%%%%%%%%%%%%%%%%%%%%%%%%%%%%%%%%%%%%%%%%%%%%%%%%%%%%%%%%%%%%%%%%%%%
\section{Color-Decorated (A)dS$_3$ Gravity: Metric Formulation}
\label{sec: metric}
So far, we described the theory in terms of the gauge field $\cA$\,,
so the fact that we are dealing with color-decorated gravity is not tangible. For the sake of concreteness and the advantage of intuitiveness, we shall recast the theory in metric formulation.

We first need to solve the torsionless conditions. This is  technically a cumbersome step. Here, we take a short way out from this problem. The idea is that, instead of solving the torsionless conditions for all the colored fields, we shall do it only for the singlet graviton, which we identified above with the metric. This will still allow us to write the action in a metric form but, apart from the gravity, all other colored fields
will be still described by a first-order Lagrangian.

In three dimensions, any spectrum with spin greater than zero can be written as 
a first-order Lagrangian which describes only one helicity mode. If one solves the torsionless conditions for the remaining non-gravity fields, the two fields describing helicity positive and negative modes will combine to generate a single field with a standard second-order Lagrangian. However, this last step appears not necessary and even impossible for certain spectra.

In the following, we will derive the full metric action for the first-order Lagrangian description. For
the second-order Lagrangian description, we shall only identify the potential term, leaving aside the explicit form of kinetic terms.

%%%%%%%%%%%%%%%%%%%%%%%%%%%%%%%%%%%%%%%%%
\subsection{Colored Gravity  around  Singlet Vacuum}

Starting from the Chern-Simons formulation, described in terms of $e^{a}$, $\o^{ab}+\O^{ab}/\ell$, 
$(\bm A,\tilde{\bm A})$ and
$(\bm\varphi,\tilde{\bm\varphi})$, we construct a metric formulation by solving the torsionless condition of the gravity sector. 
This condition is given by
\be
	d e^{a}+\left(\o^{ab}+\frac1\ell\,\O^{ab}\right)\wedge e_{b}
	+\frac{\sqrt{\s}}{N\,\ell}\,\epsilon^{abc}\,
	\tr\left(\bm{\varphi}_b\wedge \bm{\varphi}_c
	-\tilde{\bm \varphi}_b\wedge\tilde{\bm \varphi}_c\right)=0\,,
\ee
where we require $\o^{ab}$ to satisfy the standard torsionless condition \eqref{Torsion}\,. This forces $\O^{ab}=\epsilon^{abc}\,\O_{c}$ to satisfy
\be
	\O^{[a}\wedge e^{b]}
	-\frac{\sqrt{\s}}{N}\,
	\tr\left(\bm{\varphi}^a\wedge \bm{\varphi}^b
	-\tilde{\bm \varphi}^a\wedge\tilde{\bm \varphi}^b\right)=0\,.
	\label{Omega}
\ee
With the above condition together with the standard torsionless condition \eqref{Torsion}\, ,
the action \eqref{CS} can be recast to the sum of three parts:
\be
	S=S_{\rm Gravity}+
	S_{\rm CS}
	+S_{\rm Matter}\,.
	\label{S decomp}
\ee
The first term $S_{\rm Gravity}$ is the action for the (A)dS$_3$ gravity, given by\footnote{In our normalization, $d^{3}x\sqrt{|g|}=\frac16\,\epsilon_{abc}\,e^{a}\wedge e^{b}\wedge e^{c}$ .} 
\ba
	S_{\rm Gravity}[g] \eq
	\frac{\k\,N}{4\pi\,\ell}\,
	\int \epsilon_{abc}\,e^{a}\wedge
	\left(d\o^{bc}+\o^{bd}\wedge\o_{d}{}^{c}+
	\frac{\s}{3\,\ell^{2}}\,e^{b}\wedge e^{c}\right)\nn
	\eq\frac1{16\pi\,G}\int d^{3}x\sqrt{|g|}\left(R+\frac{2\,\s}{\ell^{2}}\right),
\ea
where the Chern-Simons level is related to the Newton's constant $G$, the (A)dS$_3$ radius $\ell$ and the rank of the color algebra $N$ by
\be
	\k=\frac{\ell}{4\,N\,G}\,.
\ee

The second term $S_{\rm CS}$ is the Chern-Simons action for $\mathfrak{su}(N) \oplus \mathfrak{su} (N)$ gauge algebra:
\be
	S_{\rm CS}[\bm A \tilde{\bm A}] =\frac{\k\,\sqrt{\s}}{2\pi}\int 
	\left[ \tr
	\left(\bm A\wedge d\bm A+\frac23\,\bm A\wedge 
	\bm A\wedge \bm A\right)
	-\tr\left(
	\tilde{\bm A}\wedge d\tilde{\bm A}+\frac23\,\tilde{\bm A}\wedge 
	\tilde {\bm A}\wedge \tilde {\bm A}\right)
\right]
.
	\label{CS A A}
\ee
In the uncolored Chern-Simons gravity, it is unclear whether the Chern-Simons level $\k$ has to be quantized
since the gauge group is not compact.
However, in the case of colored Chern-Simons gravity,
the level $\k$ should take an integer value 
for the consistency of $S_{\rm CS}$ \eqref{CS A A}
under a large $SU(N) \times SU(N)$ gauge transformation.

The last term $S_{\rm Matter}$
is the action for the colored massless spin-two fields
$\bm\varphi^{a}$ and $\tilde{\bm\varphi}^{a}$\,.
To derive it, we use the decompositions \eqref{B detail} and \eqref{C detail}, and simplify by using \eqref{Omega}. 
We get
\be
	S_{\rm Matter} [\bm \varphi, \tilde{\bm \varphi}] =
	\frac{1}{16\,\pi\,G}\int 
\left[	\frac{1}{N}\,L[\bm\varphi,\tilde{\bm\varphi},\ell]
	-\frac1{\ell^{2}}\,\epsilon_{abc}\,e^{a}\wedge\O^{b} (\bm \varphi, \tilde{\bm \varphi}) \wedge
	\O^{c} (\bm \varphi, \tilde{\bm \varphi}) \right]
\,,
	\label{CM inter}
\ee
where the three-form Lagrangian
$L[\bm\varphi,\tilde{\bm\varphi};\ell]$ is given by
\ba
	&&L[\bm\varphi,\tilde{\bm\varphi},\ell]
	= L_{+}[\bm\varphi,\ell]-L_{-}[\tilde{\bm\varphi},\ell]\,,\\
	\label{L ml}
	&&L_{\pm}[\bm\varphi,\ell]=2\,\sqrt{\s}\,\tr\left[
	\frac1\ell\,\bm\varphi_{a}\wedge D\,\bm\varphi^{a}
	+\frac1{\ell^{2}}\,
	\epsilon_{abc}
	\left(\frac{\pm1}{\sqrt{\s}}\, e^{a}\wedge\bm\varphi^{b}\wedge\bm\varphi^{c}
	+\frac{2}{3}\,\bm\varphi^{a}
	\wedge\bm\varphi^{b}
	\wedge\bm\varphi^{c}\right)\right].\nonumber
\ea
In this expression, the covariant derivative $D$ is with respect to both the Lorentz transformation and the $\mathfrak{su}(N)$
gauge transformation:
\bea
	D\,\bm\varphi^{a}
	&=& d\,\bm\varphi^{a}+\o^{ab}\wedge\bm\varphi_{b}
	+\bm A\wedge\bm\varphi^{a}
	+\bm\varphi^{a}\wedge\bm A\,, \nn
D\,\tilde{\bm\varphi}^{a}
	&=& d\,\tilde{\bm\varphi}^{a}+\o^{ab}\wedge \tilde{\bm \varphi}_{b}
	+\tilde{\bm A} \wedge\tilde{\bm\varphi}^{a}
	+\tilde{\bm\varphi}^{a}\wedge\tilde{\bm A}\,.
	\label{covar}
\eea
The last term in \eqref{CM inter} is an implicit function of $\bm\varphi^{a}$ and $\tilde{\bm\varphi}^{a}$\,.
It is proportional to
\be
	\epsilon_{abc}\,e^{a}\wedge\O^{b}\wedge\O^{c}
	=\frac13\,\epsilon_{abc}\,e^{a}\wedge e^{b}\wedge e^{c}\,
	\O_{[d}{}^{,d}\,\O_{e]}{}^{,e}\,,
	\label{O 2}
\ee
where $\O^{a}=\O_{b}{}^{a}\,e^{b}$\,.
% Since we solve the torsionless condition, dreibeins are the complete basis of 1-forms.
From \eqref{Omega}, they are determined to be
\be
	\O_{a}{}^{b}
	=\frac{1}{N}\,W_{a}{}^{b}(\bm\varphi,\tilde{\bm\varphi}) = \O_{a}{}^{b} (\bm \varphi, \tilde{\bm \varphi}) \,,
\ee
where $W_{a}{}^{b}(\bm\varphi,\tilde{\bm\varphi})$\, is given by
\ba	
	&& W_{a}{}^{b}(\bm\varphi,\tilde{\bm\varphi})=
	W_{a}{}^{b}(\bm\varphi)-W_{a}{}^{b}(\tilde{\bm\varphi})\,,\nn
	&&
	W_{a}{}^{b}(\bm\varphi)=4\,\sqrt{\s}\,\tr\left(
	\bm{\varphi}_{[a}{}^{b}\,\bm{\varphi}_{c]}{}^{c}
	-\frac14\,\delta^{b}_{a}\,\bm{\varphi}_{[c}{}^{c}\,\bm{\varphi}_{d]}{}^{d}
	\right).
	\label{W exp}
\ea
Here, $\bm\varphi_{b}{}^{a}$ are the components of $\bm\varphi^{a}$\,:
$\bm\varphi^{a}=\bm\varphi_{b}{}^{a}\,e^{b}$\,.
Notice that only the term \eqref{O 2}
--- which is quartic in $\bm\varphi^{a}$ and $\tilde{\bm\varphi}^{a}$ --- gives the cross couplings between $\bm \varphi$'s and $\tilde{\bm\varphi}$'s.

\subsection{First-order Description}

Gathering all above results and replacing the dreibein $e^{a}$ in terms of the metric $g_{\m\n}$\,, 
the colored gravity action reads
\be
	S
	=S_{\rm CS}
	+\frac{1}{16\pi\,G}\int \rmd^{3}x\sqrt{|g|}\left[
	R-V(\bm\varphi,\tilde{\bm\varphi})+
	\frac{2\sqrt{\s}}{N\,\ell}\,\epsilon^{\m\n\r}\,\tr\left(
	\bm\varphi_{\m}{}^{\l}D_{\n}\bm\varphi_{\r\l}
	-\tilde{\bm\varphi}_{\m}{}^{\l}D_{\n}\tilde{\bm\varphi}_{\r\l}\right)
	\right],
\label{final action}
\ee
where the covariant derivative is given by
\be
	D_{\m}\bm\varphi_{\n\r}
	=\nabla_{\m}\bm\varphi_{\n\r}
	+[\bm A_{\mu},\bm\varphi_{\n\r}]\,
\ee
and the scalar potential function is given by
\ba
	&& V(\bm\varphi,\tilde{\bm\varphi}) \nn
	= && -\frac1{N\,\ell^{2}}\,\tr\,\Big[2\,\s\,\bm I
	+4\left(\bm\varphi_{[\m}{}^{\m}\,\bm\varphi_{\n]}{}^{\n}
	+\tilde{\bm\varphi}_{[\m}{}^{\m}\,\tilde{\bm\varphi}_{\n]}{}^{\n}
	\right)
	+8\sqrt{\s}
	\left(\bm\varphi_{[\m}{}^{\m}\,\bm\varphi_{\n}{}^{\n}\,\bm\varphi_{\r]}{}^{\r}
	-\tilde{\bm\varphi}_{[\m}{}^{\m}\,\tilde{\bm\varphi}_{\n}{}^{\n}\,
	\tilde{\bm\varphi}_{\r]}{}^{\r}\right)\Big]
	\nn
	&&-\,\frac{16\,\s}{N^{2}\,\ell^{2}}\,
	\tr\Big(\bm{\varphi}_{[\m}{}^{\n}\,\bm{\varphi}_{\r]}{}^{\r}-
	\tilde{\bm \varphi}_{[\m}{}^{\n}\,\tilde{\bm \varphi}_{\r]}{}^{\r}\Big)\,
	\tr\Big(\bm{\varphi}_{[\n}{}^{\m}\,\bm{\varphi}_{\l]}{}^{\l}-
	\tilde{\bm \varphi}_{[\n}{}^{\m}\,\tilde{\bm \varphi}_{\l]}{}^{\l}\Big)
	\nn
	&&+\,\frac{6\,\s}{N^{2}\,\ell^{2}}\,
	\Big[
	\tr\left(\bm{\varphi}_{[\m}{}^{\m}\,\bm{\varphi}_{\n]}{}^{\n}-
	\tilde{\bm \varphi}_{[\m}{}^{\m}\,\tilde{\bm \varphi}_{\n]}{}^{\n}\right)
	\Big]^2\,.
	\label{V pot}
\ea
The scalar potential function consists of single-trace and double-trace parts. 
The single-trace part originates from the Chern-Simons cubic interaction, while the double-trace part originates from solving the torsionless conditions. For a general configuration, all terms in the potential 
have the same order in large $N$  as the other terms in (\ref{final action}). 
 
Already at this stage, the content of the colored gravity is clearly demonstrated: it is a theory of 
colored massless left-moving and right-moving spin-two fields, as seen from the kinetic term in (\ref{L ml})  or (\ref{final action}).
They interact covariantly with the color singlet gravity
and also with the Chern-Simons color gauge fields.
Moreover, they interact with each other through the potential $V(\bm\varphi,\tilde{\bm\varphi})$\,.
The self-interaction is governed by the constant $1/N$. The single-trace cubic interaction is stronger than the gravitational cubic interaction by the factor of $\sqrt{N}$\,. Therefore, at large $N$ and for fixed Newton's constant, the colored massless spin-two fields will be strongly coupled to each other.

%%%%%%%%%%%%%%%%%%%%%%%%%%%%%%%%%%%%%%%%%%%%%%%%%%%%%%%%%
\subsection{Second-order Description}
In principle, we could also solve the torsionless condition for the colored spin-two fields and obtain a second-order Lagrangian
(although this spoils the minimal interactions to the $\mathfrak{su}(N)$
gauge fields $\bm A$ and $\tilde{\bm A}$).
It amounts to taking linear combinations
\be
	\bm\chi_{\m\n}=\sqrt{\s}
	\left(\bm\varphi_{\m\n}-\tilde{\bm\varphi}_{\m\n}\right),
	\qquad
	\bm\t_{\m\n}=\bm\varphi_{\m\n}+\tilde{\bm\varphi}_{\m\n}\,,
\ee
and integrating out the \emph{torsion} part $\bm\t_{\m\n}$, while   keeping $\bm\chi_{\m\n}$\,. 
The resulting action is given by
\be
	S
	=S_{\rm CS}
	+\frac{1}{16\pi\,G}\int d^{3}x\sqrt{|g|}\,\big[\,
	R-V(\bm\chi)+
	\cL_{\rm CM}(\bm\chi,\nabla\bm\chi,\bm A,\tilde{\bm A})\,
	\big]\,. 
	\label{chi action}
\ee
The Lagrangian $\cL_{\rm CM}$
reads
\be
	\cL_{\rm CM}(\bm\chi,\nabla\bm\chi,\bm A, \tilde{\bm A})=\frac{1}{N}\,\tr\left(2\,
	\bm\chi_{\m\n}\,\nabla^{2}\,\bm\chi^{\m\n}
	+\cdots\right),
\ee
where the ellipses include 
other tensor contractions together with higher-order terms 
of the form, $\bm\chi^{n}\,(\nabla\,\bm\chi)^{2}$ with $n\ge1$ as well as couplings to the gauge fields $\bm A$ and $\tilde{\bm A}$. We do not attempt to obtain the complete structure of these derivative terms. 

The potential function $V(\bm\chi)$  corresponds to the extremum of 
\ba
	V(\bm\chi,\bm\t)
	\eq -\frac{2\,\s}{N\,\ell^{2}}\,\tr\Big(\bm I+
	\bm\chi_{[\m}{}^{\m}\,\bm\chi_{\n]}{}^{\n}
	+\s\,\bm\t_{[\m}{}^{\m}\,\bm\t_{\n]}{}^{\n}
	+\bm\chi_{[\m}{}^{\m}\,\bm\chi_{\n}{}^{\n}\,\bm\chi_{\r]}{}^{\r}
	+3\,\s\,\bm\chi_{[\m}{}^{\m}\,\bm\t_{\n}{}^{\n}\,\bm\t_{\r]}{}^{\r}\Big)
	\nn
	&&-\,\frac{4}{N^{2}\,\ell^{2}}\,
	\tr\Big(\bm{\chi}_{[\m}{}^{\n}\,\bm{\tau}_{\r]}{}^{\r}+\bm{\tau}_{[\m}{}^{\n}\,\bm{\chi}_{\r]}{}^{\r}\Big)\,
	\tr\Big(\bm{\chi}_{[\n}{}^{\m}\,\bm{\tau}_{\l]}{}^{\l}+\bm{\tau}_{[\n}{}^{\m}\,\bm{\chi}_{\l]}{}^{\l}\Big)
	\nn
	&&+\,\frac{6}{N^{2}\,\ell^{2}}\,
	\left[\tr\left(\bm{\chi}_{[\m}{}^{\m}\,\bm{\tau}_{\n]}{}^{\n}\right)\right]^2\,,
	\label{V x t}
\ea
along the $\bm\t_{\m\n}$ direction. As the extremum equation for $\bm\t_{\m\n}$ is linear in $\bm\t_{\m\n}$,
\bea
{\cal M}(\bm{\chi})\cdot {\bm \tau} = 0 \, , 
\label{M}
\eea
it must be that the unique solution is $\bm \tau_{\m \n} = 0$
for a generic configuration of $\bm\chi_{\m\n}$.\footnote{There can also exist nontrivial $\bm \t_{\m \n}$ solutions at special values of $\bm\chi_{\m\n}$, corresponding to kernel of ${\cal M}$ in (\ref{M}). They break the parity symmetry spontaneously, and hence of special interest. We relegate complete classification of these null solutions in a separate paper.}
Proceeding with this situation, we end up with the cubic potential for $\bm \chi_{\m \n}$\,:
\be
	V(\bm\chi)
	=-\frac{2\,\s}{N\,\ell^{2}}\,\tr\Big(\bm I+
	\bm\chi_{[\m}{}^{\m}\,\bm\chi_{\n]}{}^{\n}
	+\bm\chi_{[\m}{}^{\m}\,\bm\chi_{\n}{}^{\n}\,\bm\chi_{\r]}{}^{\r}
	\Big)\,.
	\label{V chi}
\ee
This potential has a noticeably simple form, but also has rich implications as we shall discuss in the next sections.

%%%%%%%%%%%%%%%%%%%%%%%%%%%%%%%%%%%%%%%%%%%%%%%%%%%%%%%
\section{Classical Vacua of Colored Gravity}
\label{sec: vacua}

\subsection{Identification of Vacuum Solutions}

Having identified the action in metric formulation,   
we now search for classical vacua that solve the field equations of motion:
\ba
	-\frac{\delta \cL_{\rm CM}}{\delta g_{\m\n}}
	= G_{\m\n}-\frac{1}2V(\bm\chi)\,g_{\m\n}\,,&
	\qquad
	&\frac{\delta \cL_{\rm CM}}{\delta \bm\chi_{\m\n}}
	= \frac{\partial V(\bm\chi)}{\partial \bm\chi_{\m\n}}\,,\\
	-\frac{N}{2\,\sqrt{\s}\,\ell}\frac{\delta \cL_{\rm CM}}{\delta \bm A_{\m}}
	= \epsilon^{\m\n\r}\,\bm F_{\n\r}\,,&
	\qquad
	&-\frac{N}{2\,\sqrt{\s}\,\ell}\frac{\delta \cL_{\rm CM}}{\delta \tilde{\bm A}_{\m}}
	= \epsilon^{\m\n\r}\,\tilde{\bm F}_{\n\r}\,.	
	\label{A eq}
\ea
In order to find their solutions, we assume that
the colored massless spin-two fields are covariantly constant with the trivial $\mathfrak{su}(N)$ gauge connection,
\be
	\bm A=0\, , \qquad \tilde{\bm A} = 0\,,
	\qquad\nabla_{\r}\,\bm\chi_{\m\n}=0\,.
	\label{A0}
\ee
This can be satisfied by
\be
	\bm\chi_{\m\n}
	=g_{\m\n}\,\bm X\, \qquad \mbox{for} \qquad \bm X = \mbox{constant} \, \in \, \mathfrak{su}(N)\,.
	\label{gX}
\ee
Physically, we interpret this as the colored spin-two matter acting as Higgs field. In Poincar\'e invairant field theory, the vacuum is Poincar\'e invariant, so only a scalar field $\varphi$ (which is proportional to an identity operator $\varphi \, \propto \, \mathbb{I}$) can take a vacuum expectation value, $\langle \varphi \rangle=v$\,. On the other hand, fields with nonzero spin cannot develop a nonzero expectation value since it is incompatible with the Lorentz invariance. In generally covariant field theory, where the background metric $g_{\m\n}$ plays the role of first fundamental form, the spin-2 field $\bm\chi_{\m\n}$  can similarly develop a nonzero vacuum expectation value $\langle \chi_{\m\n}\rangle=v\,g_{\m\n}$\, proportional to the metric $g_{\m\n}$, while all other fields of higher spin cannot.  We thus refer this phenomenon to as `gravitational Higgs mechanism'.

\begin{figure}[!h] 
\centering
\includegraphics[scale=0.5]{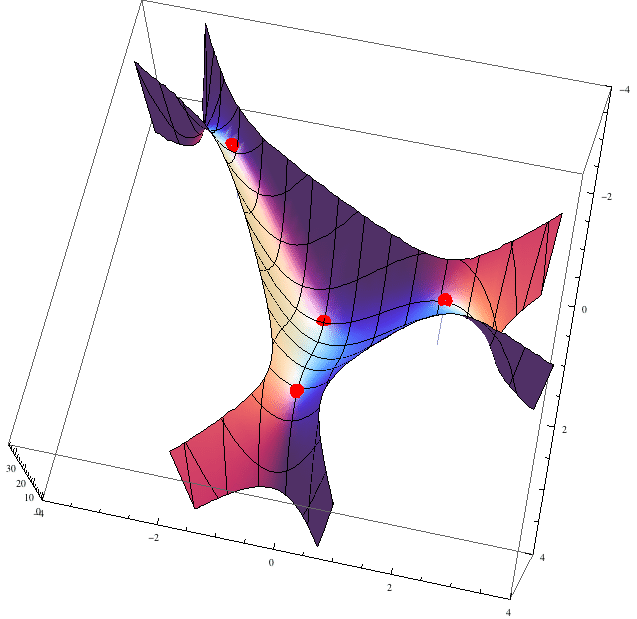} 
\caption{\sl The shape of the potential function for $\mathfrak{su}(3)$.}
\label{3d Potential}
\end{figure}

With \eqref{A0} and \eqref{gX}, the equations in the second line \eqref{A eq} trivialize and the rest reduce to
\be
	G_{\m\n}-\frac{1}2V(\bm X)\,g_{\m\n}=0\,
	\qquad \mbox{and} \qquad
	\frac{\partial V(\bm X)}{\partial \bm X}=0\,,
	\label{G V}
\ee
where $V(\bm X)=V(\bm\chi_{\m\n}=g_{\m\n}\,\bm X)$
is given by
\be
	V(\bm X)
	=-\frac{2\,\s}{N\,\ell^{2}}\,\tr\left(\bm I+3\,\bm X^{2}+\bm X^{3}\right).
	\label{V x}
\ee
From \eqref{G V}, the extremum of the potential
defines the corresponding cosmological constant:
\be
	\L=\frac{1}2\,V(\bm X)\,.
	\label{L V}
\ee
Although cubic, being a matrix-valued function,
the potential $V(\bm X)$ may admit a large number of nontrivial extrema that depends on the color algebra $\mathfrak{su}(N)$\,.
If exists, each of such extrema 
will define a distinct vacuum with a different cosmological constant (\ref{L V}).
As an illustration of this potential, consider 
the function $f(\bm X)=\frac1{N}\,\tr\left(\bm I+3\,\bm X^{2}+\bm X^{3}\right)$
for the $\bm X$ belonging to $\mathfrak{su}(3)$\,.
The $3\times3$ matrix $\bm X$ can be diagonalized by a $SU(3)$ rotation to  
\be
	\bm{X}=a\,
	{\footnotesize\begin{bmatrix}1&0&0\\0&1&0\\0&0&-2 \end{bmatrix}}
	+b\,
	{\footnotesize \begin{bmatrix}-2&0&0\\0&1&0\\0&0&1 \end{bmatrix}}. 
\ee
We plot the function $f(a,b)$ in Fig.\ref{3d Potential}. It clearly exhibits four extremum points: $(0,0)$, $(2,0)$, $(0,2)$ and $(-2,-2)$\,. 
The first point at the origin gives $f=1$,
whereas the other three points all
give $f=9$\,. In fact, these three points are connected by $SU(3)$ rotation. 

We now explicitly identify the extrema of potential function \eqref{V x} for arbitrary value of $N$\,.
The extremum points are defined by the equation:
\be
	\delta V(\bm X)
	=-\frac{6\,\s}{N\,\ell^{2}}\,\tr\left[ (2\,\bm X+\bm X^{2})\,\delta\bm X\right]=0\,.
\ee
Since $\bm X$ is traceless, it  follows that $\delta \bm X$ is also traceless. Thus, the equation reads
\be
	2\,\bm X+\bm X^{2}=
	\frac1N\,\tr\left(2\,\bm X+\bm X^{2}\right)\bm I\,.
	\label{X eq}
\ee
Since $\tr\,(\bm I+\bm X)^{2}\neq 0$ ---
otherwise it would follow from \eqref{X eq} that 
the matrix $\bm I+\bm X$ is nilpotent while having
a non-trivial trace ---
one can redefine the matrix $\bm X$ in terms of $\bm Z$\,:  
\be
	\bm Z={\sqrt{\frac{N}{\tr(\bm I+\bm X)^{2}}}}\ 
	(\bm I+\bm X)\,,
\ee
or equivalently,
\be
	\bm X=\frac{N}{\tr(\bm Z)}\,\bm Z-\bm I\,.
	\label{Z X}					
\ee
This simplifies the equation \eqref{X eq} as
\be
	\bm Z^{2}=\bm I\,.
\ee
Complete solutions of this equation, up to $SU(N)$ rotations,
are given by
\be
	\bm Z_{k}=\begin{bmatrix}
  \bm I_{\sst (N-k)\times(N-k)}   & 0  \\
    0 & -\bm I_{\sst k\times k} 
\end{bmatrix}\,,
\qquad k=0,1,\ldots, \left[\tfrac{N-1}2\right].
\label{Z(k)}
\ee
where the upper bound of $k$ is fixed by $[\frac{N-1}2]$ due to the property that 
$\bm X_{N-k}$ is a $SU(N)$ rotation of $\bm X_{k}$\,.
Notice also that, when $N$ is even, $k=\frac N2$ is excluded since it leads to $\tr(\bm Z)=0$
for which $\bm X$ is ill-defined.
Plugging the solutions \eqref{Z(k)} to the potential, we 
can identify the values of the potential at the extrema as
\be
	V(\bm X_{k})
	=-\frac{2\,\s}{\ell^{2}} \left(\frac{N}{\tr\left(\bm Z_{k}\right)}\right)^{2}
	=-\frac{2\,\s}{\ell^{2}}\left(\frac{N}{N-2k}\right)^{2}.
\ee
These values play the role of the cosmological constant at the $k$-th extremum, according to \eqref{L V}.

Let us discuss more on the potential \eqref{V x}. 
Firstly, the cubic form shows that  the potential is not bounded from below or above.
Secondly, the overall factor $\s$ shows that the overall sign of the potential depends whether we consider AdS$_3$ or dS$_3$ background.
Thirdly, we can understand better the stability of the extrema we found by considering the second variation of the potential,
\be
	\delta^{2} V(\bm X_{k})=-\frac{12\,\s}{(N-2k)\,\ell^{2}}\,
	\tr\left(\bm Z_{k}\,\d \bm X^{2}\right).
	\label{saddle}
\ee
The Hessian is not positive(or negative)-definite for an arbitrary $\d\bm X$ except the singlet vacuum $k=0$\,. So, all $k\neq 0$ vacua are saddle points
and the $k=0$ vacuum is the minimum/maximum in dS$_3$/AdS$_3$ space.
\begin{figure}[!h] 
\centering
{\includegraphics[scale=1.0]{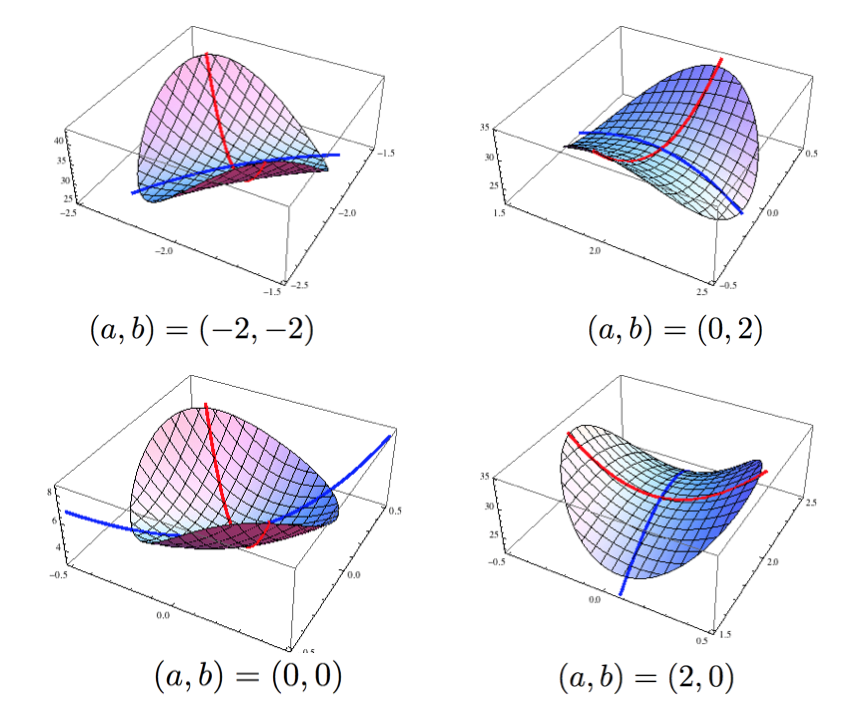}} \quad \quad
\caption{\footnotesize{\sl Potentials around each rainbow vacua with 
$N=3$\,. The $(a,b)=(0,0)$ vacuum is the minimum/maximum in dS$_3$/AdS$_3$ space. 
The others, connected by $SU(3)$ transformation, are all saddle points.
}}
\end{figure}

\subsection{$N=3$ Example and Linearized Spectrum}
\label{sec: N3}

In the standard Higgs mechansim,
 the gauge fields  combine with the Goldstone bosons 
to become massive vector fields.
In the following, we will analyze the analogous mechanism in our model of
colored gravity.
For the concreteness, let us consider the $k=1$ vacuum  solution \eqref{Z(k)} in $N=3$ case. This solution has a non-zero background for the colored matter fields
which breaks the $SU(3)$ symmetry down to $SU(2)\times U(1)$\,.
We linearize the colored matter fields $(\bm\varphi,\tilde{\bm\varphi})$
around this vacuum as
\be
	{\bm\varphi}_{\m\n}=\frac{{\bm X}_1}{2\,\sqrt{\s}}\, g_{\m\n}
	+{\bm\varphi}_{\m\n}^{\rm\sst fluc}\,,
	\qquad
	\tilde{\bm\varphi}_{\m\n}=-\frac{{\bm X}_1}{2\,\sqrt{\s}}\, g_{\m\n}
	+\tilde{\bm\varphi}_{\m\n}^{\rm\sst fluc}\,,
\ee
where the background value of $(\bm\varphi,\tilde{\bm\varphi})$
is proportional to the matrix ${\bm X}_1$ \eqref{Z X}, 
whose explicit form reads
\be
	\bm{X}_1=
	{\footnotesize\begin{bmatrix}2&0&0\\0&2&0\\0&0&-4\end{bmatrix}}\,.
\ee
The fluctuation parts of the colored matter fields and the spin-one Cherns-Simons gauge fields can be decomposed as
\ba
	{\bm\varphi}_{\m\n}^{\rm\sst fluc}\eq 
	\frac{3}{\sqrt{2}}\,\varrho^a_{\m\n}\,\bm T^a_{\mathfrak{su}(2)}	
	+\sqrt{\frac32}
	\left(\tilde\psi_{\m\n}-2\,\psi_{\m\n}\right) \bm T_{\mathfrak{u}(1)}
	+\frac{3}{\sqrt{2}}\,\phi^i_{\m\n}\,\bm T^i_{\rm\sst BS}
	\,,\nn
	\tilde{\bm\varphi}_{\m\n}^{\rm\sst fluc} \eq \frac{3}{\sqrt{2}}\,\,\tilde\varrho^a_{\m\n}\,\bm T^a_{\mathfrak{su}(2)}
	+\sqrt{\frac32}\left(\psi_{\m\n}-2\,\tilde\psi_{\m\n}\right)\bm T_{\mathfrak{u}(1)}
	+\frac{3}{\sqrt{2}}\,\tilde\phi^i_{\m\n}\,\bm T^i_{\rm\sst BS}	\,,\label{lin k1}\\
	{\bm A}_{\m}\eq 
	A^a_{\m} \,\bm T^a_{\mathfrak{su}(2)}	
	+A_\m\, \bm T_{\mathfrak{u}(1)}
	+\frac{1}{\sqrt{8}}\,A_\m^i\,{\bm Z}_1\,\bm T^i_{\rm\sst BS}\,,\nn
	\tilde{\bm A}_{\m}\eq 
	\tilde{A}^a_{\m} \,\bm T^a_{\mathfrak{su}(2)}	
	+\tilde{A}_\m\, \bm T_{\mathfrak{u}(1)}
	+\frac{1}{\sqrt{8}}\,\tilde{A}_\m^i\,{\bm Z}_1\,\bm T^i_{\rm\sst BS}\,,\label{lin k2}
\ea
in terms of the  $SU(3)$  generators:
\ba
	&\bm{T}_{\mathfrak{su}(2)}^a=
	{\footnotesize\frac{1}{\sqrt{2}}\,\begin{bmatrix}\s^a&0\\0&0\end{bmatrix}}\,,\quad
	\bm{T}_{\mathfrak{u}(1)}=
	{\footnotesize \frac{1}{\sqrt{6}}\begin{bmatrix}1&0&0\\0&1&0\\0&0&-2 \end{bmatrix}}\,,\\
	&\bm{T}_{\rm\sst BS}^1=
	{\footnotesize\frac{1}{\sqrt{2}}\,\begin{bmatrix}0&0&1\\0&0&0\\1&0&0\end{bmatrix}}\,,\quad
	\bm{T}_{\rm\sst BS}^2=
	{\footnotesize\frac{1}{\sqrt{2}}\,\begin{bmatrix}0&0&-i\\0&0&0\\i&0&0\end{bmatrix}}\,,\quad
	\bm{T}_{\rm\sst BS}^3=
	{\footnotesize\frac{1}{\sqrt{2}}\,\begin{bmatrix}0&0&0\\0&0&1\\0&1&0\end{bmatrix}}\,,\quad
	\bm{T}_{\rm\sst BS}^4=
	{\footnotesize\frac{1}{\sqrt{2}}\,\begin{bmatrix}0&0&0\\0&0&-i\\0&i&0\end{bmatrix}}\,,\nonumber
\ea
where $\s^{1},\s^2,\s^3$ are the Pauli matrices. 
Various factors in \eqref{lin k1} and \eqref{lin k2} have been introduced for latter convenience.
By plugging \eqref{lin k1}
into the original action \eqref{final action} and 
expanding the action up to quadratic order in the fluctuations, we 
obtain the perturbative Lagrangian around the $k=1$ vacuum.
We first expand the potential as
\be
	V(\bm\varphi,\tilde{\bm\varphi})
	=-\frac{2}{\ell_1^2}\left(
	\s+
	\varrho^a_{[\m}{}^{\mu}\,\varrho^a_{\n]}{}^{\n}
	+\tilde\varrho^a_{[\m}{}^{\mu}\,\tilde \varrho^a_{\n]}{}^{\n}
	-\psi_{[\m}{}^{\mu}\,\psi_{\n]}{}^{\n}
	-\tilde\psi_{[\m}{}^{\mu}\,\tilde\psi_{\n]}{}^{\n}\right)
	+\mathcal{O}(\Phi^3)\,,
	\label{pot}
\ee
and the kinetic part as
\ba
	&&\frac{2\sqrt{\s}}{9\,\ell_1}\,\epsilon^{\m\n\r}\,\tr\left(
	\bm\varphi_{\m}{}^{\l}D_{\n}\bm\varphi_{\r\l}
	-\tilde{\bm\varphi}_{\m}{}^{\l}D_{\n}\tilde{\bm\varphi}_{\r\l}\right)\label{kin}\\
	&&
	=
	\frac{\sqrt{\s}}{\ell_1}\,\epsilon^{\m\n\r}\left(
	\varphi'^a_{\m}{}^{\l}\nabla_{\n}\varphi'^a_{\r\l}
	+	\psi_{\m}{}^{\l}\nabla_{\n}\psi_{\r\l}
	+	\phi^i_{\m}{}^{\l}\nabla_{\n}\phi^i_{\r\l}
	+\frac{1}{\sqrt\s}\,\phi^i_{\m\n}\,A^i_\r\right)+c.c+\mathcal{O}(\Phi^3)\,.\nonumber
\ea
Here the $\ell_1=\ell/3$ is the radius of the $k=1$ (A)dS solution,
and $\mathcal{O}(\Phi^3)$ means the cubic-order terms in the fluctuation fields.
Combining \eqref{pot} and \eqref{kin}, 
the colored gravity action \eqref{final action} becomes
\be
	S=S_{\rm CS}
	+\frac{1}{16\pi\,G}\int d^3\sqrt{|g|}\left(R+\frac{2\,\s}{\ell^2_1}
	+\cL_{\rm\sst RS}+\cL_{\rm\sst BS}+\mathcal{O}(\Phi^3)\right),
\ee
where the Lagragian for the residual symmetry part is given by
\ba
	\ell_1^2\,\cL_{\rm\sst RS}
	\eq 
	\sqrt{\s}\,\ell_1\,\epsilon^{\m\n\r}
	\varrho^a_{\m}{}^{\l}\nabla_{\n}\varrho^a_{\r\l}
	+2\,	\varrho^a_{[\m}{}^{\mu}\,\varrho^a_{\n]}{}^{\n}+c.c.\nn
	&&
	+\,\sqrt{\s}\,\ell_1\,\epsilon^{\m\n\r}\,
		\psi_{\m}{}^{\l}\nabla_{\n}\psi_{\r\l}
		-2\,\psi_{[\m}{}^{\mu}\,\psi_{\n]}{}^{\n}
	+c.c.
	\label{RS L}
\ea
and that for the broken symmetry part by
\be
	\ell_1^2\,\cL_{\rm\sst BS}
	=
	\sqrt{\s}\,\ell_1\,\epsilon^{\m\n\r}\left(
	\phi^i_{\m}{}^{\l}\nabla_{\n}\phi^i_{\r\l}
	+\frac{1}{\sqrt\s}\,\phi^i_{\m\n}\,A^i_\r\right)
	+c.c.
	\label{BS L}
\ee 
Several remarks are in order:
\begin{itemize}

\item
In the Lagrangian $\cL_{\rm\sst RS}$ \eqref{RS L},
the fields $\varrho^a_{\m\n}$ --- associated with 
the $\mathfrak{su}(2)$ generators ---
describe the standard massless spin-two fields.
On the contrary, the field $\psi_{\m\n}$ --- associated with the
$\mathfrak{u}(1)$ generator --- describes a \emph{ghost} massless spin-two 
 due to the sign flip of the no-derivative term (why this sign  
determines whether the spectrum is ghost or not is 
explained  in Appendix \ref{sec: PM}).

\item
In the Lagrangian $\cL_{\sst\rm BS}$ \eqref{BS L} --- associated with the broken part of the symmetry ---
has an unusual  cross term with the spin-one Cherns-Simons gauge field $A^i_\mu$\,.
In fact, $A^i_\mu$ behaves as a Stueckelberg field hence can be
removed by a spin-two gauge transformation. Let us remark that this gauge choice 
is analogous to the unitary gauge in the standard Higgs mechanism. As a result, the Chern-Simons action $S_{\rm\sst CS}$ reduces from $SU(3)$ to $SU(2)\times U(1)$\,,
and the field $\phi_{\m\n}^i$ inherits the gauge symmetries of $A^i_\mu$ 
as a second-derivative form:
\be 
	\delta\,\phi^i_{\m\n}= \left(\nabla_\mu\,\nabla_\nu-\frac{\s}{\ell_1^2}\, g_{\mu\nu}\right) \xi^i\,.
\ee
This spectrum clearly combines  the massless spin-two mode with the spin-one mode in an irreducible manner.
It actually corresponds to so-called partially-massless spin-two field \cite{Deser-Waldron}.
Since our system is after all a Chern-Simons theory, there is no propagating
DoF such as a scalar field. Hence, it is clear that we cannot have a massive spin-two
as a result of symmetry breaking because it would require
not only spin-one but also a scalar mode.
We postpone more detailed analysis to the next section.

\end{itemize}

%%%%%%%%%%%%%%%%%%%%%%%%%%%%%%%%%%%%%%%%%%%%%%%%%%%%

\section{Colored Gravity around Rainbow Vacua}
\label{sec: rainbow}

We learned that there are $[\frac{N+1}2]$ many distinct vacua having different  cosmological constants. 
In this section, we study the colored gravity around each of these vacua and analyze the 
spectrum.
In principle, we can proceed in the same way as we did for the $N=3$ example in
Section \ref{sec: N3},
but there is  a more systematic way relying on the Chern-Simons formulation.

\subsection{Decomposition of Algebra Revisited}

For an efficient treatment of the colored gravity at each distinct vacuum in the Chern-Simons formulation, it is important to identify the proper decomposition of the algebra 
\eqref{g=b+c}.
For that, we revisit the isometry and the color algebra decompositions. The isometry algebra can be divided into the rotation part $\cM$ and the translation 
part $\cP$ as
\be
	\mathfrak{iso}=\cM\oplus\cP\,, 
\ee
the same as the trivial vacuum. 
For the color algebra, each vacuum spontaneously breaks the Chan-Paton $\mathfrak{su}(N)$ gauge symmetry down to $\mathfrak{su}(N-k)\oplus \mathfrak{su}(k)\oplus \mathfrak{u}(1)$, and hence the original algebra admits the decomposition:
\be
	\mathfrak{su}(N)\simeq
	\mathfrak{su}(N-k)\oplus
	\mathfrak{su}(k)\oplus \mathfrak{u}(1)\oplus \mathfrak{bs}\,. 
	\label{bs}
\ee
Here, $\mathfrak{bs}$ is the vector space 
corresponding to the \emph{broken symmetry}, spanned by $2k(N-k)$ generators. 
It is important to note that each part 
commutes or anti-commutes with the background matrix $\bm Z_{k}$ (\ref{Z(k)}) as
\be
	\big[\,\bm Z_{k}\,,\,\mathfrak{su}(N-k)\oplus
	\mathfrak{su}(k)\oplus \mathfrak{u}(1)\,\big]=0\,,
	\qquad
	\big\{\,\bm Z_{k}\,,\,\mathfrak{bs}\,\big\}=0\,.
\ee

We now decompose the entire algebra \eqref{g} according to 
\eqref{g=b+c} in terms of 
the gravity plus gauge sector $\mathfrak{b}$ and 
the matter sector $\mathfrak{c}$\,.
The former has again two parts similarly to the singlet vacuum case as
$\mathfrak{b}=\mathfrak{b}_{\rm GR}
	\oplus \mathfrak{b}_{\rm Gauge}$\,,
but the algebras to which the gravity and the gauge sectors correspond
differ from \eqref{b content}. They are
\be
	\mathfrak{b}_{\rm GR}=
	\big(\cM\otimes{\bm I}\big)
	\oplus
	\big(\cP\otimes \bm Z_{k}\big)\,,
	\qquad
	\mathfrak{b}_{\rm Gauge}=
	{\rm id}\otimes 
	\Big(\mathfrak{su}(N-k)\oplus\mathfrak{su}(k)\oplus \mathfrak{u}(1)\Big)\,.
	\label{b gr gauge}
\ee
The gauge sector is concerned only with  the unbroken part of  the color algebra. 
The algebra of the gravity sector is deformed by $\bm Z_{k}$\,, but  still satisfies the same commutation relations with
the generators:
\be
	\bm M_{ab}=M_{ab}\,\bm I\,,
	\qquad
	\bm P_{a}=P_{a}\,\bm Z_{k}\,.
	\label{mat M P}
\ee
The one-form gauge  fields associated with these sectors are given correspondingly by
\ba
	\cB_{\rm GR}\eq
	\frac12\left(\o^{ab}+\O^{ab}\right)\bm M_{ab}
	+\frac1{\ell_{k}}\,e^{a}\,\bm P_{a}\,,\nn
	\cB_{\rm Gauge}\eq
	\bm A_{+}+\bm A_{-}
	+\tilde{\bm A}_{+}+
	\tilde{\bm A}_{-}+\big(A+\tilde A\big)\,\bm Y_{k}\,,
	\label{B k}
\ea
where the $k$-vacuum radius $\ell_{k}$ is related to the singlet one as
\be
	\ell_{k} :=\left( \frac{N-2k}{N} \right) \,\ell\,,
\label{k-vacuum length}
\ee
and $\bm Y_{k}$ is  the traceless matrix:  
\be
	\bm Y_{k}=\frac{k\,\bm I_{+}
	-(N-k)\,\bm I_{-}}N\,.
\ee
Here again, the spin connection $\o^{ab}$ is the standard 
one satisfying \eqref{Torsion},
whereas $\O^{ab}$ will be determined in terms of other fields from the torsionless conditions.
The gauge fields $\bm A_{\pm}$ and $\tilde{\bm A}_{\pm}$ take values 
in $\mathfrak{su}(N-k)$ for the subscript $+$ and 
$\mathfrak{su}(k)$ for the subscript $-$\,, whereas $A$ and $\tilde A$ are
Abelian gauge fields taking values in $\mathfrak{u}(1)$\,.

In the case of non-singlet vacua,
the matter sector space has two parts:
\be
	\mathfrak{c}
	=\mathfrak{c}_{\rm CM}\oplus
	\mathfrak{c}_{\rm BS}\,.
	\label{c k}
\ee
For the introduction of each elements,
let us first define
the generators of $\mathfrak{gl}_{2}\oplus\mathfrak{gl}_{2}$ deformed by $\bm{Z}_{k}$ as
\ba
	&\bm J_{a}=J_{a}\,\bm I_{+}+\tilde J_{a}\,\bm I_{-}\,,
	\qquad
	&\bm J=J\,\bm I_{+}+\tilde J\,\bm I_{-}\,,\nn
	&\tilde{\bm J}_{a}
	=J_{a}\,\bm I_{-}+\tilde J_{a}\,\bm I_{+}\,,\qquad
	&\tilde{\bm J}=J\,\bm I_{-}+\tilde J\,\bm I_{+}\,,
	\label{twisted J}
\ea
where $\bm I_{\pm}$ are the identities associated with
$\mathfrak{u}(N-k)$ and $\mathfrak{u}(k)$\,, respectively:
\be
	\bm I_{\pm}=\frac12\left(\bm I\pm\bm Z_{k}\right).
\ee
These deformed $\mathfrak{gl}_{2}\oplus\mathfrak{gl}_{2}$
generators satisfy also the same relation as \eqref{gl2},
and they are related to $\bm M_{ab}$ and $\bm P_{a}$ \eqref{mat M P}
analogously to \eqref{MP J} by
\be
	\bm M_{ab}=\frac12\,\epsilon_{ab}{}^{c}\,
	\big(\bm J_{c}+\tilde{\bm J}_{c}\big)\,
	\qquad \mbox{and} \qquad
	\bm P_{a}=\frac{1}{2\sqrt{\s}}\,
	\big(\bm J_{a}-\tilde{\bm J}_{a}\big)\,.
\ee
Therefore, if we define the matter fields using $\bm J_{a}$
and $\tilde{\bm J}_{a}$\,, then they will
have the standard interactions with the gravity.

We now introduce each elements of \eqref{c k}.
The first one $\mathfrak{c}_{\rm CM}$ is the residual color symmetry:
\be
	\mathfrak{c}_{\rm CM}= \mathfrak{iso}\otimes 
	\Big(\mathfrak{su}(N-k)\oplus \mathfrak{su}(k)\oplus \mathfrak{u}(1)\Big)\,,
\ee
describing  colored spin-two fields
associated with the one form 
\be
	\cC_{\rm CM}=\frac1{\ell_{k}}
	\left[\left(\bm\varphi^{a}_{+}+\bm\varphi^{a}_{-}\right)
	\bm J_{a}
	+\left(\tilde{\bm\varphi}^{a}_{+}+\tilde{\bm\varphi}^{a}_{-}\right)
	\tilde{\bm J}_{a}+
	\left(\psi^{a}\,\bm J_{a}
	+\tilde\psi^{a}\,\tilde{\bm J}_{a}\right) \bm Y_{k}\,\bm Z_{k}\right].
	\label{C CM k}
\ee
The fields $\bm\varphi_{+}^{a}$ and $\tilde{\bm\varphi}_{+}^{a}$
take values in $\mathfrak{su}(N-k)$, whereas
$\bm\varphi_{-}^{a}$ and $\tilde{\bm\varphi}_{-}^{a}$
in $\mathfrak{su}(k)$\,, both transforming in the adjoint representations.
The fields $\psi^a$ and $\tilde \psi^a$ are charged under $\mathfrak{u}(1)$\,.
The matrix factor $\bm Y_{k}\,\bm Z_{k}$ is inserted to ensure
$\tr(\mathfrak{b}_{\rm GR}\,\mathfrak{c}_{\rm CM})=0$, equivalently,
\be
	\tr\big(\bm J\,\bm Y_{k}\,\bm Z_{k}\big)
	=0
	=\tr\big(\tilde{\bm J}\,\bm Y_{k}\,\bm Z_{k}\big)\,.
\ee 

The second element $\mathfrak{c}_{\rm BS}$ is what corresponds to
the broken part of the color symmetries:
\be
	\mathfrak{c}_{\rm BS}=({\rm id}\oplus \mathfrak{iso})
	\otimes \mathfrak{bs}\,.
\ee
Unlike the fields in $\cC_{\rm CM}$\,, this part does not describe massless spin-two fields. Rather, it describes
so-called partially-massless spin-two fields \cite{Deser-Waldron}, as
we shall demonstrate in the following.
The corresponding one form is given by
\be
	\cC_{\rm BS}=\frac1{\ell_{k}}\left(\bm\phi\,\bm J
	+\bm\phi^{a}\,\bm J_{a}
	+\tilde{\bm\phi}\,\tilde{\bm J}
	+\tilde{\bm\phi}^{a}\,\tilde{\bm J}_{a}\right),
	\label{C PM}
\ee
where the fields $\bm\phi^{a}$, $\bm\phi$, $\tilde{\bm\phi}^{a}$
and $\tilde{\bm\phi}$  take values in $\mathfrak{bs}$,
carrying the bi-fundamental representations of $\mathfrak{su}(N-k)$
and $\mathfrak{su}(k)$, as well as the representation of $\mathfrak{u}(1)$\,.
Because these fields anti-commute with $\bm Z_{k}$\,,
they also intertwine the left-moving and the right-moving $\mathfrak{gl}_{2}$'s. 
For instance,
\be
	\bm\phi^{a}\,\bm J_{b}=\tilde{\bm J}_{b}\,\bm\phi^{a}\,.
\ee
As a consequence, they transform differently
under Hermitian conjugate:
\be
	(\bm\phi,\bm\phi^{a},\tilde {\bm \phi},\tilde{\bm\phi}^{a})^{\dagger}
	=\left\{
	\begin{array}{c}
	(-\tilde {\bm \phi},\tilde{\bm\phi}^{a},-\bm\phi,\bm\phi^{a})
	\qquad  [\s=+1] \vspace{5pt}\\
	(-\bm\phi,\bm\phi^{a},-\tilde {\bm \phi},\tilde{\bm\phi}^{a})\qquad  [\s=-1]
	\end{array}\right.,
	\label{pm hermicity}
\ee
compared to the massless ones \eqref{varphi dagger}.

%%%%%%%%%%%%%%%%%%%%%%%%%%%%%%%%%%%%%%%%%%%%%%%%%%%%%%%%%%
\subsection{Colored Gravity around Non-Singlet Vacua}
\label{sec: 4.2}

With the precise form of the  fields 
\eqref{B k}, \eqref{C CM k}, \eqref{C PM},
we now rewrite the Chern-Simons action into a metric form. It is given by the sum of three terms as in \eqref{S decomp}. Firstly, we have the standard gravity action
\be
	S_{\rm Gravity}
	=\frac1{16\pi\,G}\int d^{3}x\sqrt{|g|}\left(R+\frac{2\,\s}{\ell_{k}^{2}}\right),
	\label{GR k}
\ee
with a $k$-dependent  cosmological constant, set by (\ref{k-vacuum length}). 
The Chern-Simons action $S_{\rm CS}$
for the gauge fields $\bm A_{+}$ for $\mathfrak{su}(N-k)$\,,
 $\bm A_{-}$ for $\mathfrak{su}(k)$ 
 and $A$ for $\mathfrak{u}(1)$ are given 
analogously to \eqref{CS A A}.
Finally, the action for the matter sector takes the following form:
\ba
	S_{\rm Matter}\eq
	\frac{1}{16\,\pi\,G}\int 
	\frac{1}{N-2k}\left(
	L[\bm\varphi_{+},\tilde{\bm\varphi}_{+},\ell_{k}]
	-L[\bm\varphi_{-},\tilde{\bm\varphi}_{-},\ell_{k}]
	+L_{\rm\sst BS}[\bm\phi,\tilde{\bm\phi},\ell_{k}]+L_{\rm cross}\right)\nn
	&&\hspace{45pt}
	-\,\frac{k(N-k)}{N^{2}}\,L[\psi,\tilde{\psi},\ell_{k}]
	-\frac1{\ell_{k}^{2}}\,\epsilon_{abc}\,e^{a}\wedge\O^{b}\wedge\O^{c}\,,  
	\label{Matter k}
\ea
where  $L$ is the massless Lagrangian given in \eqref{L ml}  
whereas $L_{\rm\sst BS}$ is  given by
\be
	L_{\rm\sst BS}[\bm\phi,\tilde{\bm\phi},\ell]
	=\frac{4\,\sqrt{\s}}{\ell}\,\tr
	\left[\left\{\tilde{\bm\phi}\wedge 
	\left(D\,\bm\phi-\frac1{\sqrt{\s}\,\ell}\,e^{a}\wedge\bm\phi_{a}\right)
	-\tilde{\bm\phi}_{a}\wedge 
	\left(D\,\bm\phi^{a}-\frac1{\sqrt{\s}\,\ell}\,e^{a}\wedge\bm\phi\right)
	\right\} \bm Z_{k}
	\right].
	\label{L pm}
\ee
The covariant derivatives $D\bm \phi^{a}$ and 
$D\bm \phi$ are given by
\ba
	D\bm\phi^{a}\eq 
	D_{\o}\bm\phi^{a}
	+
	\left(\tilde{\bm A}_{+}+\tilde{\bm A}_{-}
	+\tilde A\,\bm Y_{k}\right)
	\wedge\bm\phi^{a}-\bm\phi^{a}\wedge 
	\left({\bm A}_{+}+{\bm A}_{-}
	+A\,\bm Y_{k}\right),
	\nn
	D\bm\phi \, \, \eq
	\,d\bm\phi \, 
	+\, \left(\tilde{\bm A}_{+} +\tilde{\bm A}_{-}
	+\tilde A\,\bm Y_{k}\right)\wedge\bm\phi
	\,\,  -\,\,  \bm\phi\wedge
		\left({\bm A}_{+}+{\bm A}_{-}
	+A\,\bm Y_{k}\right),
\ea
and similarly for the tilde counter parts.
The other terms in \eqref{Matter k} give additional interactions:
the last term gives quartic interaction through  $\O_{a}{}^{,b}$\,:
\be
	\O_{a}{}^{,b}
	=\frac{1}{N-2k}\,
	\Big[ W^{a}_{b}(\bm\varphi_{+},\tilde{\bm\varphi}_{+})
	+W^{a}_{b}(\bm\varphi_{-},\tilde{\bm\varphi}_{-})
	+W_{\rm\sst BS}{}^{a}_{b}(\bm\phi,\tilde{\bm\phi})\Big]
	-\frac1{k(N-k)}\,W^{a}_{b}(\psi,\tilde\psi)\,,
\ee
where $W^{a}_{b}$ is given by \eqref{W exp}
and $W_{\rm\sst BS}{}^{a}_{b}$ by
\be
	W_{\rm\sst BS}{}^{a}_{b}(\bm\phi,\tilde{\bm\phi})=
	8\,\sqrt{\s}\,\tr\left[\left(
	\tilde{\bm{\phi}}_{[a}{}^{[b}\,\bm{\phi}_{c]}{}^{c]}
	-\frac14\,\delta^{b}_{a}\,\tilde{\bm{\phi}}_{[c}{}^{c}\,\bm{\phi}_{d]}{}^{d}
	\right) \bm Z_{k}
	\right].
\ee
The term $L_{\rm cross}$, given by
\ba
	L_{\rm cross}\eq
	\frac{4\sqrt{\s}}{\ell_{k}^{2}}
	\,\tr\Big[
	\left(\bm\varphi_{+}^{a}-\bm\varphi_{-}^{a}+\psi^{a}\,\bm Y_{k}\right)
	\wedge
	\left(\tilde{\bm\phi}
	\wedge\bm\phi_{a}
	+\tilde{\bm\phi}_{a}\wedge\bm\phi\right)+\nn
	&&\hspace{40pt}+\,\epsilon_{abc}
	\left(\bm\varphi^{a}_{+}\wedge\bm\varphi^{b}_{+}
	+\bm\varphi^{a}_{-}\wedge\bm\varphi^{b}_{-}
	\right)\wedge\psi^{c}\,\bm Y_{k}\,\bm Z_{k}\Big]
	-\left([\ \, ]\leftrightarrow\widetilde{[\, \ ]}\right),
\ea
is the cross terms originating from the Chern-Simons cubic interactions.

In principle, we can further simplify the action as we did in the singlet
vacuum case.
However, already at this level, we can extract a lot of physics.
\begin{itemize}
\item
We have a scalar potential as a function 
of four fields $\bm\varphi_{\pm}$, $\psi$, $\bm\phi$ (and their tilde counter parts)
and the point where all fields vanish correspond to the extremum point
whose potential value gives the cosmological constant $-\s/\ell_{k}^{2}$\,.
This potential should be a shift of the  potential 
$V(\bm\varphi,\tilde{\bm\varphi})$
\eqref{V pot}
defined around the singlet vacuum,
hence it will admit all other vacua as extrema.

\item
The interaction strength for each field can be 
easily read off from the action. 
The gravity and gauge interaction have the same strength 
controlled by $G$ and $\k$ as in the singlet vacuum case.
The interaction of colored spin two fields $\bm\varphi_{\pm}$  
is weakened --- the coefficient changed from $N$ to $N-2k$\,.
The same for the broken-symmetry field
$\bm\phi$\,.
Finally,  $\psi$ has 
interaction strength controlled by $N^{2}/[k(N-k)]$. 
Therefore, when the color symmetry is maximally broken, that is $N-2k\sim1$\,,
the interaction between all these fields becomes as weak as the gravitational one.

\item 
Let us conclude this section with the summary of the field content around the $k$-vacuum.
At first, we have the graviton and $\mathfrak{su}(N-k)\oplus\mathfrak{su}(k)\oplus \mathfrak{u}(1)$ Chern-Simons gauge fields.
Next, about the colored matter fields,
there are $(N-k)^2-1$ fields for $(\bm\varphi_+,\tilde{\bm\varphi}_+)$,
$k^2-1$ for $(\bm\varphi_-,\tilde{\bm\varphi}_-)$ 
and 1 for $(\psi,\tilde\psi)$\,.
They are all massless spin-two fields, but $(\bm\varphi_-,\tilde{\bm\varphi}_-)$ and $(\psi,\tilde\psi)$
--- hence $k^2$ fields --- are in fact ghost.
For the broken symmetry part, we have
$2k(N-k)$ fields for $(\bm\phi,\tilde{\bm\phi})$\,.
The latter describes so-called partially-massless fields
and its proper analysis is the subject of the next section.

\end{itemize}

%%%%%%%%%%%%%%%%%%%%%%%%%%%%%%%%%%%%%%%%%%%%%%%%%%%%%%%

\subsection{Partially Massless Spectrum Associated with Broken Color Symmetry}

Around a non-singlet vacuum, the fields $\bm\varphi_{\pm}$
and $\psi$ both describe massless spin-two fields
having the same quadratic Lagrangian given by \eqref{L ml}.
On the other hand, the fields $\bm\phi$ corresponding to
the broken part of the color symmetries have
different quadratic Lagrangian \eqref{L pm}, hence describe
different spectrum. We have already mentioned that
they correspond to partially-massless  fields \cite{Deser-Waldron}.
In this section, we  analyze the quadratic Lagrangian \eqref{L pm} to prove this statement. Here, we concentrate on AdS$_3$. To get the dS$_3$ result, 
it is sufficient to replace $\ell$ by $i\,\ell$.

Though the Lagrangian \eqref{L pm} has a rather non-standard form
involving cross term between $\bm\phi$ and $\tilde{\bm\phi}$
together with an insertion of $\bm Z_{k}$\,, it can always be diagonalized with the help of the Hermiticity property \eqref{pm hermicity}.
Therefore, for the spectrum analysis,
it will suffice to consider $S_{\rm\sst BS}[\phi,\phi^{a}]$ taking the following expression:
\be
	S_{\rm\sst BS}[\phi,\phi^{a}]
	=\int
	\phi\wedge \left(d \phi- \frac1{\ell}\,e_{a}\wedge\phi^{a}\right)
	-\phi_{a}\wedge 
	\left(D\phi^{a}-\frac1{\ell}\,e^{a}\wedge \phi\right),
	\label{PM action}
\ee
with the AdS dreibein and spin connection $(e^{a},\o^{ab})$\,.
We first note that this action admits the gauge symmetries with parameters $(\varepsilon,\varepsilon^{a})$\,, 
\be
	\delta\, \phi=
	d\,\varepsilon-\frac{1}{\ell}\,e^a\,\varepsilon_{a}\,,
	\qquad
	\delta\, \phi^{a}=
	D\,\varepsilon^{a}-\frac{1}{\ell}\,e^a\,\varepsilon\,,	
\ee
which come from the Chern-Simons gauge symmetries.

For a closer look of this action
involving three fields $h_{\m\n}=e^{a}{}_{(\m}\,\phi_{\n)a}$\,,
$f_{\m\n}=e^{a}{}_{[\m}\,\phi_{\n] a}$
and $\phi_{\mu}=e^{a}{}_{\m}\,\phi_{a}$, 
we consider two
different but equivalent paths:
\begin{itemize}
\item
We first derive the equation of motion for one-form fields $\phi^a$ and $\phi$\,.
They are given by
\be
D\phi^a
- \frac{1}{\ell} e^a \wedge\phi
=0\,,
\qquad
d\phi
- \frac{1}{\ell}e^a\wedge \phi_a
=0\,.
\ee
The second equation implies that the antisymmetric field $f_{\mu\nu}$
is the field strength of $\phi_\m$\,:
$f_{\mu\nu}=\ell\,\partial_{[\m}\phi_{\n]}$\,.
 Then, by gauge fixing $\phi_{\mu}$ to zero with the gauge parameter $\varepsilon^{a}$, 
 the field $f_{\mu\nu}$ decouples from the first equation.
We thus end up with only one field $h_{\m\n}$
 satisfying the equation of motion, 
\be
	\nabla_{[\m}h_{\n]\r}=0\,,
	\label{PM eeq}
\ee
and the gauge symmetry,
\be
	\delta\,h_{\m\n}
	=\left(\nabla_\mu\,\nabla_\nu-\frac{1}{\ell^2}\, g_{\mu\nu}\right)\varepsilon\,\,.
	\label{PM gauge}
\ee
This coincides with the gauge symmetry 
of partially-massless spin-two field \cite{Deser-Waldron}.

\item
Instead of first deriving the equation and then gauge fixing to 
\mt{\phi_{\mu}=0}\,,
one can reverse the procedure.
We first gauge fix and eliminate $\phi_{\mu}$ field in the action and obtain
\be
	S_{\rm\sst PM}[\phi_{\m\n}]=\int d^{3}x\sqrt{|g|}\ 
	\e^{\m\n\r}\,\phi^{\l}{}_{\m}\,
	\nabla_{\n}\,\phi_{\r\l}\,,			
\ee
modulo a boundary term. We note that the field $\phi_{\m\n}$ contains both of the symmetric part $h_{\m\n}$ and the antisymmetric part 
$f_{\m\n}$.
Only $h_{\m\n}$ admits the gauge symmetries \eqref{PM gauge}. 
The equation of motion is now given by
\be
	C_{\m\n,\r}:=\nabla_{[\m}h_{\n]\r}+\nabla_{[\m}f_{\n]\r}=0\,.
	\label{PM Eq}
\ee
The totally anti-symmetric part $C_{[\mu\nu,\r]}=\partial_{[\m}f_{\n\r]}=0$
can be readily solved as $f_{\m\n}=\partial_{[\m}\,a_{\n]}$\,.
With the field redefinition $h_{\m\n}\to h_{\m\n}-\nabla_{(\m}a_{\n)}$\,,
the trace of the above equation,
$C^{\r}{}_{\m,\r}=0$\,, gives
\be
a_\m=\frac{\ell^2}{2}\left(\nabla^\r h_{\m\r}-\nabla_\m {h^\r}_\r
\right),
\label{a as H}
\ee
Taking now a divergence of $C_{\mu\nu,\rho}$\,,
we arrive at the second-order equation,
\be\label{PM2derivative}
	\nabla^{\r}C_{\r(\m,\nu)}=
	G^{\sst\rm lin}_{\mu\nu}+
	\frac{1}{\ell^2}
	\left(h_{\mu\nu}-g_{\mu\nu}\,h^{\r}{}_{\r}\right)=0\,,
\ee
with the linearized Einstein tensor $G^{\sst\rm lin}_{\mu\nu}$\,.
One can also check that the mass 
in the above equation corresponds to that of a partially-massless field.
Furthermore, using Bianchi identity, we deduce that the left-hand side of \eqref{a as H}  vanishes, so does $f_{\m\n}$\,.
Therefore, we end up with the same equation \eqref{PM eeq}.\footnote{Strictly speaking, 
the equation \eqref{PM2derivative} alone
is weaker than the first-order one \eqref{PM eeq}. 
The former describes one propagating degrees of freedom, 
while the latter does not have any bulk mode
and corresponds to the spectrum described by \eqref{PM action}. 
Note that the latter partially-massless spectrum is 
what the three-dimensional conformal gravity contains
analogously to the four-dimensional case \cite{DJW,Joung:2015jza}. 
To recapitulate, in three dimensions (not in higher dimensions), 
there are two kinds of partially-massless fields
for the maximal depth, which includes the spin-two partially-massless spectrum.
We shall discuss more about this subtlety in the companion paper \cite{Gwak:2015jdo}.}

\end{itemize}

\section{Discussions}
\label{sec: discussion}

In this paper, we proposed a Chan-Paton color-decorated 
gravity in three dimensions and studied its properties. 
We have shown that the theory describes a gravitational system of colored massless spin-two matter fields coupled to $\mathfrak{su}(N)$ gauge fields.
These matter fields have a non-trivial potential 
whose extrema have $[\frac{N+1}2]$ different values of cosmological constant. 
All the extremum points but the origin spontaneously break the $\mathfrak{su}(N)$ color symmetry 
down to $\mathfrak{su}(N-k)\oplus \mathfrak{su}(k)\oplus \mathfrak{u}(1)$\,.
We found that the spin-two Goldstone modes corresponding to the broken part of the symmetries are combined with the gauge fields and become partially-massless spin-two fields. In the vacua with large $k\sim N/2$\,, 
the interactions of the matter fields are
as weak as the gravitational one. In the small $k$ vacua, their interaction becomes strong by the factor of  $\sqrt{N}$\,.

\begin{figure}[!h] 
\centering
{\includegraphics[scale=0.45]{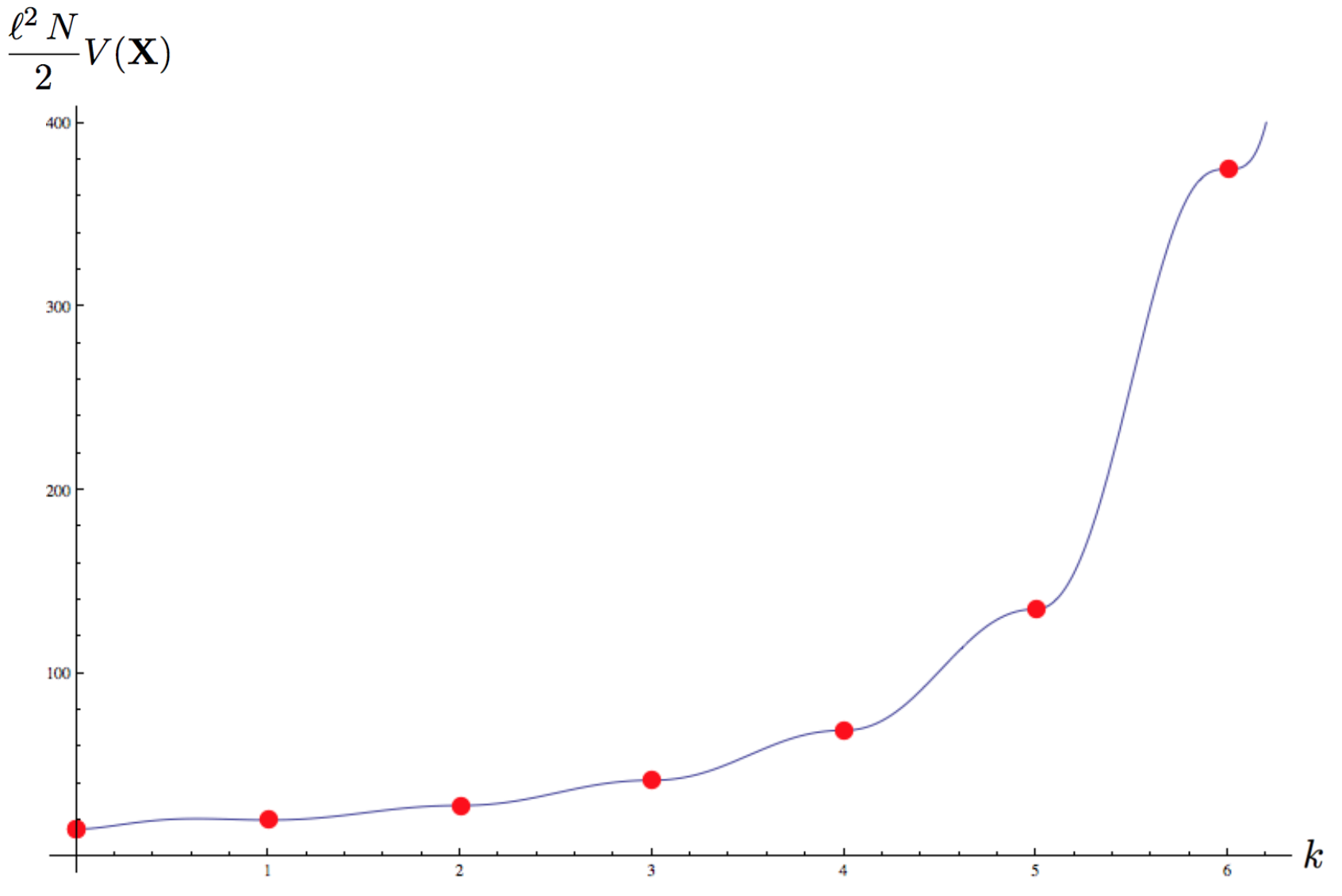}} \quad \quad \quad \quad
\caption{Potential of the colored gravity in dS ($N=15$):
$k$ is the parameter of a curve in 
$\mathfrak{su}(15)$ that passes through the extremum points.}
\label{stair}
\end{figure}

Considering the dS$_3$ branch, 
the potential takes a spiral stairwell shape (Fig.\ref{stair})
with $[\frac{N+1}2]$ many steps,  having split  cosmological constants that range from $\L=1/\ell^{2}$ at the lowest step all the way up to 
$\sim N^{2}\,\L$ at the highest step. The spacing gets dense in lower steps, while sparse in higher steps. 
If such features continue to hold in higher dimensions, the colored gravity with large $N$ might be very relevant for the early universe cosmology in that the universe begins in an inflationary epoch with a large cosmological constant at a very high stairstep. The colored matter are weakly coupled there, and hence they are not confined.
As the state of the universe decays towards lower stairsteps,
the effective cosmological constant decreases sequentially and eventually exits the inflation.
The colored matter fields start to interact stronger
and eventually form heavy color-neutral composites.
It is in this synopsis that the spin-two colored matter fields might play a novel role in the current paradigm of the inflationary cosmology.

We also speculate on a novel approach to the three-dimensional quantum colored gravity. 
At large $N$, the contribution of the $\cO(N/2)$ multiple vacua in the path integral might be captured by the $\mathfrak{su}(N)$ random matrix model 
given by
\be
{\cal Z}_{\rm MM} =	\int d\bm X\,\exp\left[i\,c\,V(\bm X)\right].
\ee 
 It would be also interesting to explore ab initio definition of the three-dimensional quantum gravity starting from
  tensor-field valued matrix models.

This work brings in many open problems worth of further investigation. First of all, 
extensions to (higher-spin) supergravity as well as the analysis of the asymptotic symmetries
\cite{W alg,SA} are imminent. 
Further extensions to color-decoration of the known higher-spin gravity in three-dimensional Lifshitz spacetime \cite{Gary:2014mca}
and flat spacetime \cite{flat} are also straightforward. Extension to higher-dimensional spacetime is also highly interesting. A version of such situation was already studied in the context of AdS/CFT correspondence  \cite{Aharony:2015zea}. Vasiliev equations for color-decorated higher-spin theories needs to be better understood, along with higher-dimensional counterpart of the stairstep potential we found in three dimensions. As the color dynamics is described by Chern-Simons gauge theory, one might anticipate to formulate colored gravity in any dimensions in terms of a version of Chern-Simons formulation, perhaps, along the lines of \cite{Bars:2001ma} and \cite{Bonezzi:2015bza}.
Quantum aspects of color-decorated gravity is an avenue to be explored. In particular, consequences and implications of strong color interactions among colored spin-two fields. Turning to the inflationary cosmology, it would be interesting to understand how the color-decoration modifies the infrared dynamics of interacting massless spin-two fields at super-horizon scales. This brings one to investigate stochastic dynamics of these fields, as would be described by color-decorated version of the Langevin dynamics \cite{Starobinsky:1986fx, Rey:1986zk}.

\acknowledgments
We are grateful to Marc Henneaux, Jaewon Kim, Jihun Kim, Sasha Polyakov, Augusto Sagnotti and Misha Vasiliev for many useful discussions. 
This work was supported in part by the National Research Foundation of Korea through the grant NRF-2014R1A6A3A04056670 (SG, EJ), and the grants 2005-0093843, 2010-220-C00003 and 2012K2A1A9055280 (SG, KM, SJR). 
The work of EJ is also supported by the Russian Science Foundation grant 14-42-00047 associated with Lebedev Institute. The work of KM was supported by the BK21 Plus Program funded by the Ministry of Education(MOE, Korea) and National Research Foundation of Korea (NRF).

\appendix

\section{Spin-Two Fields in Three Dimensions}
\label{sec: PM}
In this section, we briefly review massive and (partially) massless spin-two field 
in three dimensions. 
Let us begin with the standard Fierz-Pauli massive spin-two action
in (A)dS$_{d+1}$ background,
\ba
	\cL_{\sst\rm FP}(\chi)
	\eq -\frac12\,\nabla^\r\,\chi^{\m\n}\,\nabla_\r\, \chi_{\m\n}+
	\nabla^\n\,\chi_{\m\n}\,\nabla_\r\,\chi^{\m\r}
	-\nabla_\m\,\chi^\r{}_\r\,\nabla_\n\,\chi^{\m\n}+
	\frac12\,\nabla_\m\,\chi^\r{}_\r\,\nabla^\m\,\chi^\s{}_\s\nn
	&&
	+\frac{\s}{\ell^2}\,\left(\chi^{\m\n}\,\chi_{\m\n}+\frac{d-2}2\,\chi^\m{}_\m\,\chi^\n{}_\n\right)
	-\frac12\, m^2\,\left(	\chi^{\m\n}\,\chi_{\m\n}
	-\chi^{\m}{}_{\m}\,\chi^{\n}{}_{\n}\right),
	\label{FP}
\ea
whose equations of motion reduce to the Fierz system:
\be
	\left(\nabla^2+\frac{2\,\s}{\ell^2}-m^2\right) \chi_{\m\n}=0\,,
	\qquad
	\chi^{\m}{}_{\m}=0=\nabla^\mu\,\chi_{\m\n}\,.
\ee
In terms of the lowest energy $\D$\,, the mass parameter $m$ is given as 
\be
	\s\,\ell^2\,m^2=\D\left(\D-d\right)=\mu^2-\frac{d^2}4\,,
	\qquad \mu=\D-\frac{d}2\,.
\ee
In AdS, a very massive field corresponds to a large real $\m$\,,
whereas in dS it corresponds to a  large pure imaginary $\mu$\,.
When the mass term of the action takes a special value,
the action acquires gauge symmetries:
there are two such points, $m^2=0$ for \emph{massless-ness}
 and $m^2=-\frac{d-1}{\ell^2}\,\s$ for \emph{partially-massless-ness}.
Due to the gauge symmetries, these spectra have smaller number of DoF
than the massive one.
In particular, partially-massless spin-two field has the same amount of  DoF
as massless spin-two and massless spin-one fields.
The basic properties of massless and partially-massless spectra
are summarized in Table \ref{tab PM}.
\begin{table}[h]
\centering
\begin{tabular}{|c|c|c|c|c|}
  \hline
  Spectrum & $m^2$ & $\D_+$ & $\mu$ & Gauge symmetry \\
  \hline
  Massless & 0 & $d$ & $\frac{d}2$ & $\d\,\chi_{\m\n}=\nabla_{(\m}\,\xi_{\n)}$ 
  \\
  partially-massless & $-\frac{d-1}{\ell^2}\,\s$ & $d-1$ & $\frac{d-2}2$ & 
 $\delta\,\chi_{\mu\nu}= 
  \left(\nabla_\mu\,\nabla_\nu-\frac{\s}{\ell^2}\, g_{\mu\nu}\right) \xi$ \\
  \hline
\end{tabular}
\caption{Massless and partially-massless spin-two fields}
  \label{tab PM}
\end{table}

In three dimensions, any spin-two spectrum can be described 
in terms of a first-derivative Lagrangian.
Again beginning with the massive Lagrangian \eqref{FP}, we can
reformulate the Lagrangian into
\be
\cL_{\rm\sst FD}(\chi,\t)
=\frac12\,\e^{\m\n\r}\left(\t_\m{}^\l\,\nabla_\n\,\chi_{\r\l}
+\chi_\m{}^\l\,\nabla_\n\, \t_{\r\l}\right)
+\mu\,\left(\t_{[\m}{}^\m\,\t_{\n]}{}^\n
+\frac{\s}{\ell^2}\,\chi_{[\m}{}^\m\,\chi_{\n]}{}^\n\right),
\label{First derivative massive action}
\ee
by introducing an auxiliary field $\t_{\m\n}$\,. Here, the tensors $\chi_{\m\n}$
and $\t_{\m\n}$ do not have any symmetry properties.
By integrating out $\t_{\m\n}$ --- that is by plugging in the solution
 $\t_{\m\n}(\chi)$ of its own equation --- one can show that 
the antisymmetric part of $\t_{\m\n}$ drops  and 
the Lagrangian \eqref{First derivative massive action} reproduces 
the Fierz-Pauli Lagrangian \eqref{FP} up to a factor\,:
\be	
	\cL_{\rm\sst FD}(\chi,\t(\chi))=\frac1{\m}\,\cL_{\rm\sst FP}(\chi)\,.
	\label{FD to FP}
\ee
It is more convenient to recast the Lagrangian \eqref{First derivative massive action}
in terms of  $\varphi_{\m\n}$ and $\tilde\varphi_{\m\n}$\,:
\be
	\chi_{\m\n}=\sqrt{\s}\,\ell
	\left(\varphi_{\m\n}-\tilde\varphi_{\m\n}\right),
	\qquad
	\t_{\m\n}=\varphi_{\m\n}+\tilde\varphi_{\m\n}\,.
	\label{EL equation}
\ee
so that the massive spin-two Lagrangian splits into the parity breaking spin $+2$ and 
spin $-2$ parts:
\be
\cL_{\rm\sst  FD}(\chi,\tau)
=\cL_{+2}(\varphi)+\cL_{-2}(\tilde\varphi)\,.
\ee
Here, the self-dual massive spin $\pm2$ Lagrangian \cite{sd} is given by
\be
	\cL_{\pm 2}(\varphi)
	=\pm \sqrt{\s}\,\ell\,\e^{\m\n\r}\,\varphi_\m{}^\l\,\nabla_\n\,\varphi_{\r\l}
+2\,\mu\, \varphi_{[\m}{}^\m\,\varphi_{\n]}{}^\n\,.
	\label{CS like action}
\ee
Let us remark an unusual feature of this parity breaking massive spin-two Lagrangian
in three dimensions: the sign of the \emph{mass-like} term actually determines 
whether the Lagrangian is ghost or not, the positive sign for the unitary case
and the negative sign for the ghost, whereas 
the sign of the \emph{kinetic-like} term determines the sign of the spin. These can be seen, for example, by dualizing the above first-derivative Lagrangian to the second-order Lagrangian and relating $\mu$ to the sign in front of this Lagrangian. 
For this reason, one can render a unitary spin-two field to a ghost one
by only modifying its mass-like term in the first order description of three dimensional
theories.
Throughout this paper, we encounter three different cases:
firstly, the $\mu=1$ case corresponds to unitary massless spin-two field,
whereas the $\mu=-1$ case gives ghost massless spin-two field.
The case of $\mu=0$ describes  partially-massless spin-two field,
which  does not admit any two-derivative description as is clear
from \eqref{First derivative massive action} and \eqref{FD to FP}.

\end{document}